\documentclass[a4paper,12pt]{article}

\pdfoutput=1

\usepackage{tkz-euclide}

\usepackage[colorinlistoftodos]{todonotes}

\usepackage[utf8]{inputenc}
\usepackage{float}
\usepackage{array}
\usepackage{amsmath,amsfonts}
\usepackage{graphicx,textcomp,colortbl,booktabs,enumerate,wrapfig}
\usepackage{indentfirst,tabularx,color,psfrag,hypernat}
\usepackage{authblk}

\usepackage[leftbars]{changebar}
\usepackage[leftcaption]{sidecap}
\usepackage[T1]{fontenc}
\usepackage{mathptmx}
\usepackage[utf8]{inputenc}
\usepackage{wrapfig}
\usepackage{colortbl}
\usepackage{array}
\usepackage[compact]{titlesec}
\usepackage[abs]{overpic}
\usepackage[english]{babel}
\usepackage{amsthm}

\newtheorem*{remark}{Remark}

\usepackage{colortbl}

\usepackage{array}
\newcolumntype{Y}{>{\centering\arraybackslash}X}

\usepackage{multirow}
\usepackage{setspace}
\usepackage{xcolor}
\usepackage{pst-text}

\usepackage{wrapfig}
\usepackage{colortbl}
\usepackage{array}
\usepackage[compact]{titlesec}
\usepackage[abs]{overpic}
\usepackage{tabu}
\usepackage{setspace}
\usepackage{xcolor}

\usepackage{pst-text}
\usepackage{setspace}
\usepackage{multicol}

\usepackage[flushleft]{threeparttable}

\usepackage{xcolor}
\usepackage{amsthm}
\usepackage{hyperref}
\hypersetup{
    pdffitwindow=false,
    pdfstartview={FitH},
    pdftitle={Power-Law Attenuation},
    pdfauthor={Bharat B. Tripathi},
    pdfsubject={},
    pdfcreator={Bharat B. Tripathi},
    pdfkeywords={Piecewise Parabolic Method, Shear Shock Waves, Traumatic Brain Injury, Relaxation Mechanisms}
    urlcolor=cyan
}

\providecommand{\keywords}[1]{\textbf{\textit{keywords---}} #1}

\usepackage[top=2cm,bottom=2cm,left=2cm,right=2cm]{geometry}

\newcommand{\bea}{\begin{eqnarray}}
\newcommand{\eea}{\end{eqnarray}}
\newcommand{\nbea}{\begin{eqnarray*}}
\newcommand{\neea}{\end{eqnarray*}}

\newcommand{\ba}{\begin{align}}

\newcommand{\be}{\begin{equation}}
\newcommand{\ee}{\end{equation}}
\newcommand{\ea}{\end{align}}

\newcommand{\bfmM}{{\mbox{\boldmath{$M$}}}}

\newcommand{\bfmW}{{\mbox{\boldmath{$W$}}}}

\newcommand{\etal}{{\mbox{\it et al.} }}

\definecolor{MyGreen}{HTML}{006400}
\definecolor{maroon}{HTML}{800000}

\makeatletter
\renewcommand*\env@matrix[1][*\c@MaxMatrixCols c]{%
  \hskip -\arraycolsep
  \let\@ifnextchar\new@ifnextchar
  \array{#1}}
\makeatother

\makeindex

\usepackage{colortbl}

\usepackage{xcolor}

\usepackage{subcaption}

\makeatletter
\newcommand{\printfnsymbol}[1]{%
  \textsuperscript{\@fnsymbol{#1}}%
}
\makeatother

\usepackage{tikz,xcolor}
\definecolor{lime}{HTML}{A6CE39}
\DeclareRobustCommand{\orcidicon}{
	\begin{tikzpicture}
	\draw[lime, fill=lime] (0,0) 
	circle [radius=0.16] 
	node[white] {{\fontfamily{qag}\selectfont \tiny ID}};
	\draw[white, fill=white] (-0.075,0.095) 
	circle [radius=0.007];
	\end{tikzpicture}
	\hspace{-2mm}
}
 
\foreach \x in {A, ..., Z}{
	\expandafter\xdef\csname orcid\x\endcsname{\noexpand\href{https://orcid.org/\csname orcidauthor\x\endcsname}{\noexpand\orcidicon}}
}

\begin{document}

\def\floatpagepagefraction{1}
\def\textpagefraction{.001}

\title{\bf An atlas of the heterogeneous viscoelastic brain  with local power-law attenuation synthesised using Prony-series}

\author{Oisín Morrison \orcidA}
\author{Michel Destrade \orcidB}
\author{Bharat B. Tripathi \orcidC \thanks{Corresponding author, bharat.tripathi@universityofgalway.ie}}
\affil{\normalsize School of Mathematical and Statistical Sciences, University of Galway, University Road, Galway, Ireland.} 
\date{}

\maketitle

\begin{abstract}

This review addresses the acute need to acknowledge the mechanical heterogeneity of  brain matter and to accurately calibrate its local viscoelastic material properties accordingly. 
Specifically, it is important to compile the existing and disparate literature on attenuation power laws and dispersion to make progress in wave physics of brain matter, a field of research that has the potential to explain the mechanisms at play in diffuse axonal injury and mild traumatic brain injury in general. 
Currently, viscous effects in the brain are modelled using Prony-series, i.e., a sum of decaying exponentials at different relaxation times. Here we collect and synthesise the Prony-series coefficients appearing in the literature for twelve regions: brainstem, basal ganglia, cerebellum, corona radiata, corpus callosum, cortex, dentate gyrus, hippocampus, thalamus, grey matter, white matter, homogeneous brain, and for eight different mammals: pig, rat, human, mouse, cow, sheep, monkey and dog. Using this data, we compute the fractional-exponent attenuation power laws for different tissues of the brain, the corresponding dispersion laws resulting from causality, and the averaged Prony-series coefficients.

\end{abstract}

\noindent

\maketitle

\section*{Glossary}
We adopt the following conventions in this paper:
\begin{itemize}
    \item We reserve the following meanings for line markers:
    \begin{figure}[htbp]
    \centering
        \includegraphics[trim=5 120 5 120, clip, width=0.8\textwidth]{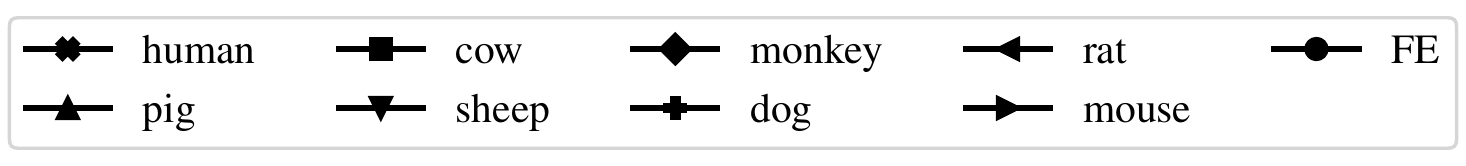}\qquad
        \caption{Legends used for different animals. {FE refers to data from finite element models.}}
    \end{figure}
    \item Owing to space limitations, it was not always possible to have legends given on all subplots. In such cases, the legends on any of the subplots apply for all of the other subplots in the figure.
    \item When error regions are shown in figures, they correspond to the region spanned by the one standard deviation errors of the parameters.
\end{itemize}

\section{Introduction}

According to the World Health Organisation, neurological disorders are one of the greatest threats to public health, with traumatic brain injury (TBI) being the leading cause of death and disability in children and young adults around the world \cite{WHO2006}. 
The problem is growing, and it is expensive as well as life-threatening. 
In the Republic of Ireland alone, a small country of five million inhabitants, about 350 million euros were spent on TBIs in 2010, out of nearly 6 billion euros spent in total on brain disorders \cite{Gustavsson2011}. 
It is thus of critical importance that a better understanding of TBI is achieved to help combat this issue. 

\begin{figure}[h!]
\begin{picture}(190,190)
\put(0,0){
\includegraphics[width=0.45\textwidth]{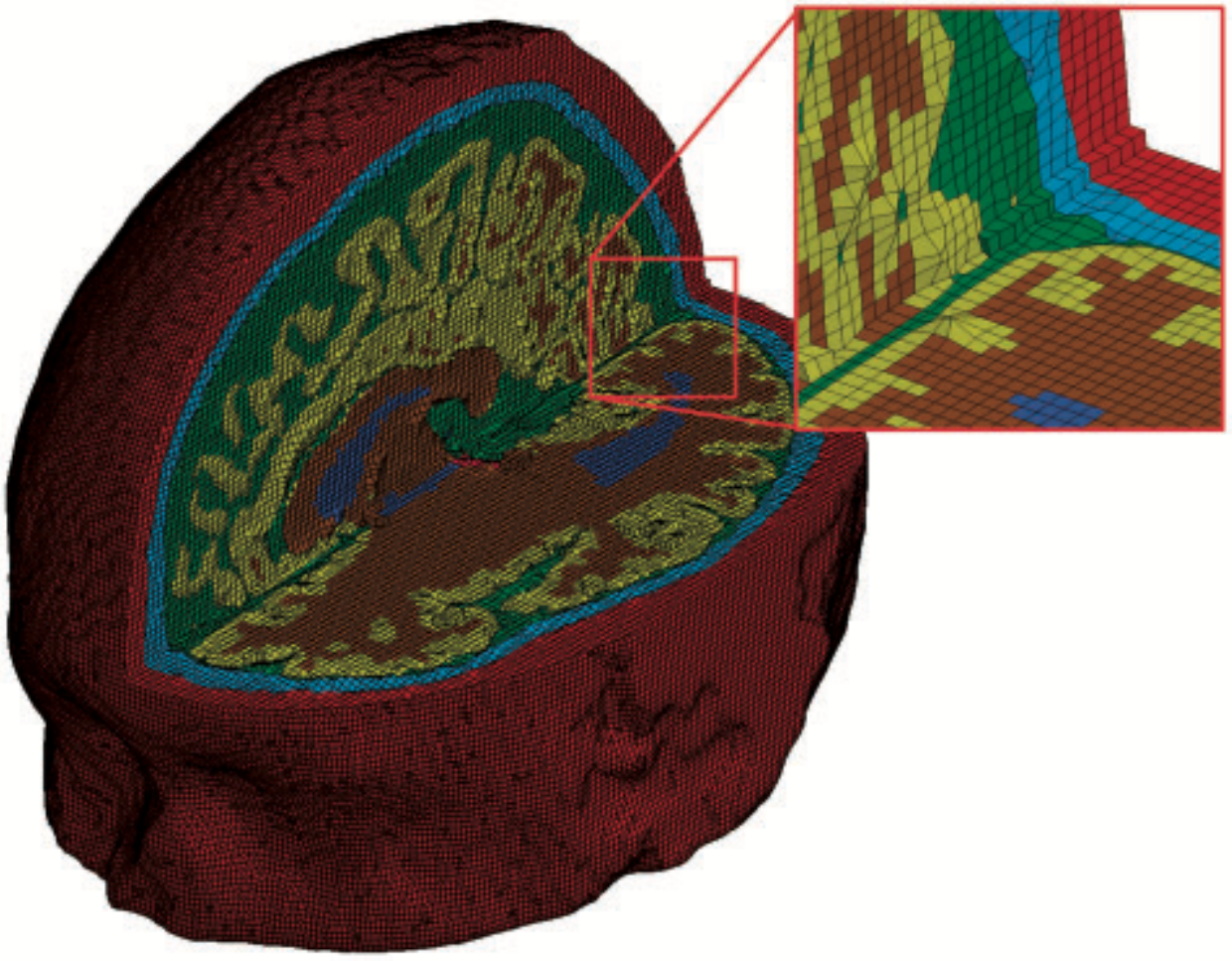}\qquad
\includegraphics[width=0.45\textwidth]{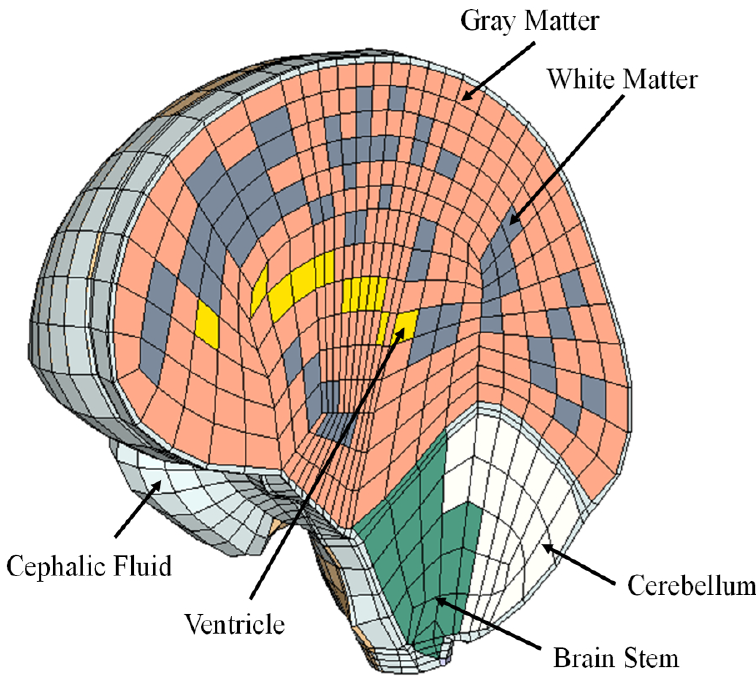}
}
\put(10,170){(a)}
\put(250,170){(b)}
\end{picture}
\caption{Two recent Finite Element (FE) head models, incorporating differing elastic properties for different areas of the brain, but the same (homogeneous) viscoelastic data everywhere. 
           (a): The finite element mesh of the high fidelity 3D model from Imperial College London \cite{ICM}.  
           Colour coding is: skin (red), skull (light blue), cerebrospinal fluid (green), grey matter (yellow), white matter (brown) and ventricles (dark blue).
           (b): The UCD Head Trauma model, originally designed by Horgan and Gilchrist \cite{UCDv1} (picture taken from Cinelli \etal \cite{Cinelli2019}). Note that its most recent version does include viscous heterogeneity \cite{UCDv2}.}
    \label{fig:FE-images}
\end{figure}

Most clinical indicators used for predicting TBI, typically linear and/or rotational accelerations, are global and not appropriate to evaluate regional brain strains and strain rates. 
But these local deformations and motions play an important role in the development of mild TBI events, such as concussion in contact sports or repetitive impacts over a lifetime \cite{Mihalik2017}.
Hence there is a pressing need for accurate material parameters that can be used in detailed finite element (FE) computer simulations \cite{Griffths2022, MacManus2022}, see Figure \ref{fig:FE-images} for two recent models.

\begin{figure}[ht!]
    \centering
        \includegraphics[width=0.9\textwidth]{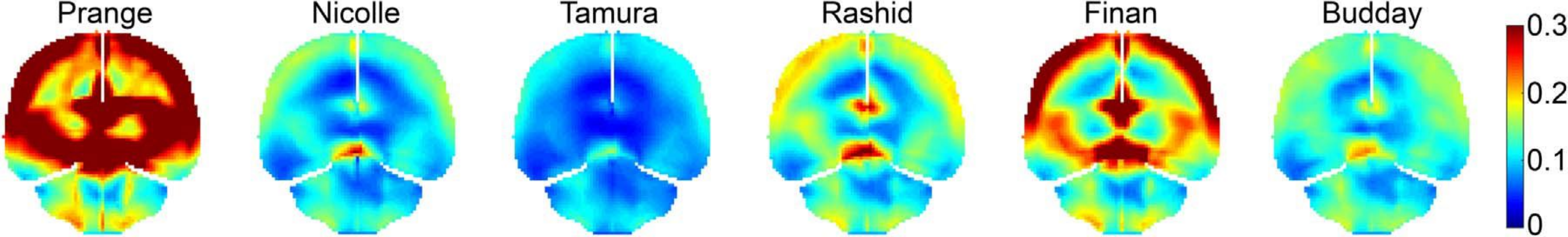}
           \caption{When the same FE model is used to simulate the same high-velocity impact, but with different material parameters from the experimental literature, it yields very different predictions of the cumulative maximum principal strains generated  \cite{Zhao2018}.
           Here the experimental data is taken (left to right) from Refs.~\cite{Prange2002, Nicolle2004, Tamura2007, Rashid2014, Finan2017, Budday2017a}
           }
    \label{fig:Zhao-et-al-18}
\end{figure}

However, there is an enormous amount of variation in the viscoelastic parameters used by existing FE models, due to dated experimental sources, differing testing protocols, temperature, type of tissue, type of animal, post-mortem times, tissue preservation modes, and many other factors. 
The brain is also often considered as a homogeneous tissue from the point of view of viscoelastic properties, while it has been experimentally observed to be heterogeneous in that respect \cite{Antonovaite2021}. 
The disparity in experimental data and the assumption of homogeneity are problematic when it comes to studying mild TBI, because they lead to very different predictions when the same event is simulated, as shown by Zhao \etal \cite{Zhao2018}, see Figure \ref{fig:Zhao-et-al-18}.

Recently shear shock waves were generated and observed experimentally in the brain, and proposed as a possible explanation for diffuse axonal injury \cite{Espindola2017a, Espindola2017b}, a major type of TBI. 
Furthermore, in direct impact injuries, it has been observed that injuries can occur far from the point of impact \cite{Graham1995}. 
The reason for this distant effect has not yet been established, but the formation of shear shock waves has been hypothesised to be a possible mechanism. 
Theoretically, cubic non-linearity must be invoked to model these nonlinear shear waves \cite{Zabolotskaya2004}; it follows that they generate mostly odd harmonics \cite{Destrade2019}, and very high local accelerations \cite{Tripathi2021}. 
Importantly, these high local accelerations are not generated instantantly instead are a result of cumulative nonlinear effect. The maximum acceleration is reached after a few centimetres of propagation in brain, before dissipating. 
Recent studies in 2D head phantoms have furthermore shown that this mechanism can predict peak accelerations far from the point of impact \cite{Tripathi2021, Chandrasekaran2022}. 
This is thus a promising and important hypothesis to test because it could have major repercussions for the prediction and understanding of TBIs, the design of helmets and other protective headgear, and the suitability of existing finite element (FE) models for modelling TBI. 
Importantly, a biofidelic modelling of the wave physics involved in shear shock wave formation and propagation requires accurate experimental data for the heterogeneous material properties of the brain -- specifically, attenuation power laws and dispersion relations.

To the best of our knowledge, there exists only one map detailing the viscoelastic properties of the heterogeneous brain, namely the recent paper by Hiscox \etal \cite{Hiscox2020}. 
In that work, the authors collect  storage and loss moduli data using a shear wave at the given frequency of 50 Hz and magnetic resonance elastography (MRE) imaging. They give this data for subcortical grey matter structures, white matter tracts, and regions of the cerebral cortex.

In this paper, we go a step further by providing viscoelastic data, used in FE models and recent experimental data, valid for multiple frequency ranges and for twelve key regions of the brain: brainstem, basal ganglia, cerebellum, corona radiata, corpus callosum, cortex, dentate gyrus, hippocampus, thalamus, grey matter and white matter, and homogeneous brain. We also provide viscoelastic data also for eight different animals: pig, rat, human, mouse, cow, sheep, monkey and dog.
We conglomerate multiple Prony-series data, the most common implementation of viscoelastic effects used in current state-of-the-art FE models. From this data, we synthesise average attenuation power laws, dispersion relations and also Prony-series.

\section{Theoretical background}

Soft solids like tissues are often modelled using hyperelastic models capable of describing large strain nonlinear deformations. At the same time, tissues are often highly attenuating and dispersing, i.e, the excitation amplitude decays with time and distance, and different frequencies travel at different speeds. Conventionally these effects are modelled using the linear \cite{Tschoegl1989} and quasi-linear \cite{Fung1993} viscoelastic theories. 

\subsection{Linear viscoelasticity} \label{sec:LV}

In linear viscoelasticity, the stress response to a constant strain decreases with time, a feature which is referred to as the stress relaxation of the material. This is modelled using the fading memory or hereditary integral:
\be
\sigma(t) = \int_{-\infty}^{t} m(t-\tau) \epsilon(\tau) d\tau := m(t)*\epsilon(t),
\label{eq:mt}
\ee
where $\sigma(t)$ is the stress (in Pa), $\epsilon(t)$ is the strain (dimensionless), and $m(t-\tau)$ is the instantaneous stress-response function to an impulse in strain $\epsilon(\tau)$ imposed at time $\tau$ for the time interval $t-\tau$. This is the so-called convolution operation, denoted by ``$*$''. 
In the frequency space, this hereditary integral can be written as 
\be
\sigma(\omega) = M(\omega)\epsilon(\omega),
\label{eq:dynamicMod}
\ee 
where $M(\omega)$ is the \emph{dynamic modulus}, corresponding to the impulse response of the material.  

However, in solid mechanics, the step-response is often more relevant than the instantaneous response. 
The memory integral can be rewritten as:
\be
\sigma(t) = \int_{-\infty}^{t} g(t-\tau) \frac{\partial \epsilon(\tau) }{\partial \tau} d\tau = g(t)*\frac{\partial \epsilon(t) }{\partial t} =
\frac{\partial g(t) }{\partial t}*\epsilon(t),
\label{eq:gt}
\ee
where $g(t)$ is the stress response to unit-step strain, often called the \textit{relaxation function}. The last equality in the above equation is due to the commutative property of the convolution integral. Also, equation \eqref{eq:mt} and equation \eqref{eq:gt} give the connection
\be
m(t) = \frac{\partial g(t) }{\partial t},
\ee
or in frequency space
\be
M(\omega) = i\omega G(\omega),
\ee
leading to   
\be
\sigma(\omega) = M(\omega)\epsilon(\omega) = i\omega G(\omega)\epsilon(\omega),
\label{eq:RelaxMod}
\ee 
where $G(\omega)$ is the complex \textit{relaxation modulus}.

Conventionally, the relaxation functions presented in the TBI literature are approximated using a \textit{Prony-series} of decreasing exponentials,  
\be
\label{eq:prony}
 g(t) =  M_{\infty} + \sum_{j=1}^{N} M_j \exp(-t/\tau_j), 
\ee
where $\tau_j = \eta_j/M_j$ ($j= 1,\ldots,N$) corresponds to the $j^{\rm th}$ Maxwell element, which is a Hookean element with elastic modulus  $M_j$ placed in series with a Newtonian element with coefficient of viscosity $\eta_j$.

Note that $g(0) = M_{\infty} +  \sum_{j=1}^{N} M_j =: M_0$, which defines the latter quantity. 
Then a dimensionless Prony-series $\hat g(t)$ with $\hat g(0)=1$ can be defined as 
\be
\label{eq:dimensionless_prony}
 \hat{g}(t) =  \hat{M}_{\infty} + \sum_{j=1}^{N} \hat{M}_j \exp(-t/\tau_j), 
\ee
where $\hat M_j = M_j/M_0$.

In frequency space, the corresponding dynamic modulus, $M(\omega)$, or relaxation modulus, $G(\omega)$,  can be written as 
\be
 M(\omega) = i \omega G(\omega) =  M_{\infty} + \sum_{j=1}^{N} M_j \frac{i \omega \tau_j}{1+ i \omega \tau_j}.
\label{eq:G-GMB-N}
\ee

Using this complex modulus, the attenuation attenuation in soft solids can be quantified using the \textit{quality factor} $Q(\omega)$ defined as \cite{OConnell1978}
\begin{equation}
\label{eq:quality_def}
Q(\omega) = \frac{\rm Re\{M(\omega)\}}{\rm Im\{M(\omega)\}} = \frac{M'(\omega)}{M''(\omega)},
\end{equation} 
where $M'(\omega) =  \text{Re}\{M(\omega)\}$ and $M''(\omega) = \text{Im}\{M(\omega)\}$ are the \textit{storage modulus} and the \textit{loss modulus},  respectively, given explicitly by
\bea
\label{eq:storage}
 M'(\omega) ={\rm Re} \{M(\omega)\} &=& M_{\infty} + \sum_{j=1}^{N} M_j\frac{\omega^2\tau_j^2}{1+ \omega^2\tau_j^2},\\
\label{eq:loss}
 M''(\omega) ={\rm Im} \{M(\omega)\} &=& \sum_{j=1}^{N} M_j\frac{\omega\tau_j}{1 + \omega^2\tau_j^2}.
\eea
Note that whilst these equations can be evaluated for any value of $\omega$, it is not physically meaningful to evaluate them over all frequencies as the Prony-series are fitted over a finite time interval. Specifically, it is valid to evaluate these functions at the (angular) frequencies $\omega = \beta_j,~j= 1,\ldots,N$, where $\beta_j = 1/\tau_j$. 
A suitable frequency range can thus be computed from the Prony-series coefficients as $[\min_j \beta_j, \max_j \beta_j]$. 
In the case of a one-term Prony-series, this would give a single point and thus then the extended frequency range $[0.1 \beta_1, 10 \beta_1]$ is used, see Nicolle \etal \cite{Nicolle2004, Nicolle2005}. This is also consistent with conventions of the commercial finite element solver Abaqus \cite{Abaqus2022}. 

The other physical behaviour associated with the attenuation is dispersion due to causality \cite{Waters2000}.
Consider the  linear shear wave equation in an elastic media,
\begin{equation}
  \frac{\partial^2 u}{\partial t^2} = \dfrac{1}{\rho}\dfrac{\partial \sigma}{\partial x} = \dfrac{\mu}{\rho}\dfrac{\partial^2 u}{\partial x^2} = c^2 \frac{\partial^2 u}{\partial x^2},
\end{equation} 
where $\rho$ is the mass density, $\mu$ is the shear modulus, and $c=\sqrt{\mu/\rho}$ is the shear wave speed.
On substituting the harmonic solution
\begin{equation}
 u = \exp[i(\omega t -kx)],
 \label{eq:harmonic}
\end{equation} 
where $u$ is the particle displacement, $\omega$ is the angular frequency, and $k$ is the wavenumber, we find the following connection for the phase velocity $c$,
\begin{equation}
 c = \omega/k. 
\label{eq:phase}
\end{equation}

Now consider the viscoelastic case as in \cite{Moczo2014}, where
\begin{equation}
  \sigma = m(t)*\epsilon = m(t) * \frac{\partial u}{\partial x},
\end{equation} 
so that the wave equation reads
\begin{equation}
  \rho \frac{\partial^2 u}{\partial t^2} =  m(t) * \frac{\partial^2 u}{\partial x^2}.
\end{equation} 
Taking the Fourier transform $\mathcal F$ with respect to time of the above equation gives
\begin{equation}
 (i\omega)^2 \mathcal{F}\{ u \}  = \frac{M(\omega)}{\rho} \frac{\partial^2 }{\partial x^2}\mathcal{F}\{ u \}.
 \label{eq:ftVE}
\end{equation} 
To calculate the right hand side of the above equation, let us rewrite equation \eqref{eq:harmonic} as
\begin{equation}
u(x,t) = \exp(i\omega t)\exp(-i K(\omega) x),
\end{equation} 
where $K(\omega)$ is the complex wavenumber in the viscoelastic media. 
Then equation \eqref{eq:ftVE} gives
\begin{equation}
 \frac{K(\omega)}{\omega} =  \sqrt{\frac{\rho}{M(\omega)}}.
\end{equation} 
Equation \eqref{eq:harmonic} can be rewritten using $K(\omega) = K'(\omega) + i K''(\omega)$ as
\begin{equation}
 u = \displaystyle \exp\{K''(\omega) x\} \exp\{i(\omega t -K'(\omega) x)\},
\end{equation} 
showing that the (real) phase velocity $c$ is given by 
\begin{equation}
\label{eq:c_omega_from_prony}
 \frac{1}{c(\omega)} = \frac{K'(\omega)}{\omega} = \text{Re}\left\{ \sqrt{\frac{\rho}{M(\omega)}} \right\}.
\end{equation}
Note $c(\omega)$ is not a Fourier transform, it is just a function in frequency space. Equation \eqref{eq:c_omega_from_prony} furthermore yields two solutions but only the principal solution is valid (the other yields $c(\omega)<0)$, which is unphysical).

The two quantities $Q(\omega)$ and $c(\omega)$ are then used to compute the \textit{attenuation} $\alpha(\omega)$ via the relation \cite{Tripathi2019, Moczo2014}:
\begin{equation}
\label{eq:quality_relation}
Q(\omega) =  \frac{1}{2}\left[\frac{\omega}{c(\omega)\alpha(\omega)} - \frac{c(\omega)\alpha(\omega)}{\omega} \right]. 
\end{equation}
On solving this quadratic equation in $\alpha(\omega)$ we get  
\bea
\alpha(\omega)=\frac{-Q + \sqrt{Q^2+1}}{{c(\omega)}/{\omega}},
\eea
while ignoring the non-physical solution where $\alpha(\omega)<0$.   

Alternatively, the attenuation of transient waves like ultrasound/shear wave in soft solids is commonly characterized using a \textit{fractional-exponent power law},
\be
\alpha(\omega) = a\omega^b = \alpha_0 f^{b},
\ee
where $a$, $b$ and $\alpha_0$ are constants.
Alternatively, in log-log space, $\ln{\alpha}$ follows an empirical linear law: $\ln{\alpha} = \ln{a}+b\ln{\omega}$.

\subsection{Quasi-Linear Viscoelasticity}\label{sec:QLV}

For large amplitude deformations, assuming a linear behaviour is no longer valid as the stress and strain exhibit a nonlinear relationship of relaxation. 
Fung \cite{Fung1993} proposed the concept of \textit{quasi-linear viscoelasticity} (QLV), with the assumption of multiplicative decomposition of the stress into a dimensionless relaxation function of time $\hat{g}(t)$ with $\hat{g}(0)=1$ and the instantaneous elastic stress $d \sigma_e(t)/dt $. On applying the superposition principle, we get 
\bea
\sigma(t) = \int_0^t \hat{g}(t-\tau) \frac{d \sigma_e(\tau)}{d \tau} d\tau
          = \int_0^t \frac{d \hat{g}(t-\tau)}{d \tau} \sigma_e(\tau) d\tau
          = \int_0^t \hat{m}(t-\tau) \sigma_e(\tau) d\tau.
\eea
The QLV formulation therefore ends up using the entire mathematical formulation of linear viscoelastic theory as described in the previous section. We can then relate $g(t)$ and $\hat{g}(t)$, as:
\bea
g(t)= M_0 \hat{g}(t)= M_\infty + \sum_{j=1}^N M_j e^{- t/\tau_j},
\eea
where $M_0$ is the instantaneous shear modulus of the hyperelastic strain energy density. \autoref{tab:M_0} gives the expressions for $M_0$ of some commonly used hyperelastic models.

\begin{table}

\begin{center}
\begin{threeparttable}
\small{
\begin{tabular}{| l | p{5.7cm} | l |  }
 \hline\\[-8pt]
 \textbf{Hyperelastic Model} & $\bfmW$ & \textbf{$\bfmM_0$} \\[2pt]
 \hline \hline\\[-8pt]
 Neo-Hookean & $\tfrac{1}{2}\mu\left(I_1-3\right)$ & $\mu$ \\[4pt]
 \hline\\[-8pt]
 Mooney-Rivlin & $C_1 \left( I_1-3\right)+ C_2 \left( I_2-3\right)$ & $2(C_1+C_2)$  \\[4pt]
 \hline\\[-10pt]
 2-term Polynomial$^a$ & $C_{10} \left( I_1-3\right)+ C_{01} \left( I_2-3\right) + C_{20} \left( I_1-3\right)^2 + C_{02} \left( I_2-3\right)^2$ & $2(C_{10}+C_{01})$  \\[4pt]
 \hline\\[-10pt]
 Ogden & $\sum_{n=1}^N \frac{\mu_n}{\alpha_n} \left( \lambda_1^{\alpha_n}+\lambda_2^{\alpha_n}+\lambda_3^{\alpha_n} -3 \right)$ & $\tfrac{1}{2} \sum_{n=1}^N \mu_n \alpha_n$ \\[4pt]
 \hline \\[-10pt]
 Gasser-Ogden-Holzapfel$^b$ & $\tfrac{1}{2}\mu \left( I_1-3 \right) + \tfrac{k_1}{2k_2} \left[e^{k_2( I_1 - 3)^2} -1\right]$ & $\mu$ \\[2pt]
 \hline

\end{tabular}}
    \begin{tablenotes}
      \small
      \item[$^a$]\hspace{-4pt}{\emph{Without cross term $C_{11} \left( I_1-3\right) \left( I_2-3\right)$}}
      \item[$^b$]\hspace{-4pt}{\emph{In the isotropic case}}
    \end{tablenotes}
  \end{threeparttable}
\end{center}

\caption{Instantaneous shear moduli of common hyperelastic strain-energy densities.}
\label{tab:M_0}
\end{table}

These two viscous modelling approaches are the two most common approaches used for describing viscoelastic effects.
Another approach is the fractional viscoelastic model \cite{Mainardi2010}, which is still not fully adopted due to its mathematical complexities.

\subsection{Calculating Prony-series from attenuation power laws}

It is possible to compute averaged attenuation power laws from the data, but many current models are heavily reliant on the use of Prony-series. As a result, it is important to provide a means of determining averaged Prony-series from an attenuation power law $\alpha(\omega)=a \omega^b$, valid over an angular frequency range $[\omega_1, \omega_2]$.

Firstly, the dispersion can be calculated for $\omega \in [\omega_1, \omega_2]$ via the Kramers-Kronig relation \cite{Waters2000}:
\bea
\frac{1}{c(\omega)}-\frac{1}{c(\omega_0)} =
\left\{ 
\begin{array}{ll}
a \tan\left(\frac{b\pi}{2}\right)(\omega^{b-1}- \omega_0^{b-1}); & {\text{when }} b \in (0,2)\backslash \{1\} \vspace{12pt}\\
-\frac{2}{\pi} a \omega_0^b (\ln\omega - \ln \omega_0);     &   b =1
\end{array}
\right. .
\label{eq:KKC} 
\eea
Note that this calculation requires a reference value $c(\omega_0)$. 
Furthermore, we note that the case $b=1$ will not occur for our fitted parameters. 

Ergo, using equation \eqref{eq:KKC} and equation \eqref{eq:quality_relation}, one can directly compute the inverse quality factor. 
Recall that the inverse of the quality factor is also directly obtainable from a Prony-series via equation \eqref{eq:quality_def}, and note that the value of $M_0$ does not influence the quality factor. 
This means that a dimensionless Prony-series with parameters $\hat{M}_j = \frac{M_j}{M_0}$ can also be used. It is thus possible to write directly:
\bea
\label{eq:curve_fitting}
Q(\omega) & =
 & \frac{1}{2}\left[\frac{\omega}{a \omega^b c(\omega_0)} \left( a c(\omega_0) \tan\left(\frac{b\pi}{2}\right)(\omega^{b-1}- \omega_0^{b-1}) + 1 \right) \right. \nonumber \\
 & & \left. \qquad - \frac{a \omega^b c(\omega_0)}{\omega} \left( a c(\omega_0) \tan\left(\frac{b\pi}{2}\right)(\omega^{b-1}- \omega_0^{b-1}) + 1 \right)^{-1} \right] \\
 & = & \displaystyle \frac{  \hat{M}_\infty+\sum_{j=1}^N \hat{M}_j \frac{\omega^2 \tau_j^2}{1+ \omega^2 \tau_j^2}  }{ \sum_{j=1}^N \hat{M}_j \frac{\omega \tau_j}{1+ \omega^2 \tau_j^2}  }. \nonumber
\eea
Thus, together with the additional constraint that $\hat{M}_\infty + \sum_{j=1}^{N} \hat{M}_j = 1$, it is possible to directly curve fit the $N$-term Prony-series once given $N$. As per Abaqus recommendations, the order of the Prony-series should not be larger than the number of logarithmic decades spanned by the test data \cite{Abaqus2022}. 
Thus, this furthermore sets $N$ as
\bea
\label{eq:order_prony}
N=\Bigl \lfloor \log_{10}  \left( \frac{\omega_2}{\omega_1} \right) \Bigl \rfloor.
\eea

Lastly, it remains to compute the value of $M_0$, which is done via the formula \cite{Tripathi2019}
\bea
\label{eq:get_M_0}
M_0 = \rho c(\omega_0)^2 \frac{| \hat{M}(\omega_0) | + \text{Re}\{ \hat{M}(\omega_0)\}}{2 | \hat{M}(\omega_0) |^2 },
\eea
where $\hat{M}(\omega)$ refers to the dynamic modulus derived using the dimensionless Prony-series.

\section{Methods}

\subsection{Overview of the literature review}

We collected a total of 181 differing Prony-series from 48 different experimental papers, spanning twelve regions of interest and eight different animal types, see in the supplementary materials. 
The cortex was the most commonly measured region in the dataset, with 43 Prony-series. The other tissues had fewer data: brainstem (23), corona radiata (19), homogeneous brain (19), cerebellum (18), hippocampus (18), corpus callosum (17), thalamus (11), dentate gyrus (7) and basal ganglia (6). 
We also made a point of collecting data on the species used in the experiments, to investigate the effects of surrogate tissues. 
The most commonly used animal surrogate was porcine tissue, with 56 Prony-series. 
A total of eight different types of animals were used in our collected experimental data - namely, rat (52), human (45), mouse (13), cow (12), sheep (1), monkey (1), and dog (1).

We only collected recent experimental data (from the past 25 years), from a variety of experimental protocols, including  indentation tests, shear tests, tensile tests and compression tests.  All of these protocols were testing ex-vivo brain tissue. 

In-vivo testing is possible by magnetic resonance elastography (MRE), but there are some limitations and assumptions associated with current MRE methods \cite{Hiscox2016, Low2016, Kalra2019}. Indeed, large discrepancies between various MRE measurements exist, sometimes by an order of magnitude \cite{Hiscox2016}. 
There are also discrepancies between the results of mechanical tests and elastography results, such as for uniaxial compression \cite{Rosen2019}. Budday et al. \cite{Budday2020} noted this discrepancy for experiments looking into age-dependence for brain tissue. There are also issues with reconstruction methodologies for MRE \cite{Babaei2021, Bilston2018}. Consequently, we did not collect MRE experimental results here, to remove this source of additional variation.

The data was averaged irrespective of experimental protocols, species, sex, temperature, or other factors. 
Ordinarily this approach might be problematic, because  it is well known that factors such as age \cite{Kalra2019, Murphy2019}, sex \cite{Arani2015, Sack2009}, animal \cite{Vink2018, Dai2018}, experimental protocol \cite{Chatelin2010, Gefen2004}, temperature \cite{Rashid2012c, Liu2017, Peters2017}, preservation \cite{Rashid2013b},  humidity \cite{Forte2017} and post-mortem time \cite{Weickenmeier2018, Finan2019} do influence experimental results for brain. 
As a result, comparison between different experimental results is difficult and studies often only compare with experiments using similar techniques \cite{Hrapko2008, Antonovaite2021}. 
Nonetheless, current FE models are using data from a range of experiments using different protocols. 
Therefore, for better understanding the data, a comparison of experimental data must be undertaken regardless of experimental protocols and other factors. Furthermore, such is the degree of variation in parameters reported in the literature, that these influences can be largely neglected in comparison. For example, Chatelin \etal \cite{Chatelin2010} found in their review that the disparity in results was independent of experimental protocol.

In relation to Prony-series used in FE models, we found a total of 31 unique Prony-series. {A total of 23 different FE models were considered in this work. In alphabetical order, they are the following: ADAPT \cite{ADAPT}, ANISO KTH v1 \cite{ANISOKTHv1}, ANISO KTH v2 \cite{ANISOKTHv2}, ATLAS \cite{ATLAS}, Cai \etal (CAI) \cite{CAI}, Chen \etal (CHEN) \cite{CHEN}, ICM \cite{ICM}, Khanuja \& Unni (KHANUJA) \cite{KHANUJA}, KTH v2 \cite{KTHv2}, SIMON v0 \cite{SIMONv0}, SIMON v1 \cite{SIMONv1}, Subramaniam \etal (SUBRAM) \cite{SUBRAM}, Tse \etal (TSE) \cite{TSE}, UCD v1 \cite{UCDv1}, UCD v2 \cite{UCDv2}, WSUBIM \cite{WSUBIM}, Yang \etal (YANG) \cite{YANG}, ULP v0 \cite{ULPv0}, ULP v1 \cite{ULPv1a, ULPv1b}, WHIM 1 \cite{WHIMv1}, WHIM 2 \cite{WHIMv2a, WHIMv2b}, Yang \etal (YANG) \cite{YANG} and YEAHM \cite{YEAHMa, YEAHMb} models}. The most commonly modelled tissue is the homogeneous brain, with 12 different Prony-series. Most data consist of only one-term Prony-series. The same problems with variations due to differing experimental protocols also apply to these datasets. Furthermore, there are also multiple instances of differing Prony-series being derived from the same experimental sources. 
This discrepancy is due to differences in fitting methods. 
In some cases, even the order of the Prony-series can change between studies - for example, from the data of Shuck and Advani \cite{Shuck1972}, the WHIM v2 model obtains a 2-term Prony-series \cite{WHIMv2a, WHIMv2b}, whilst the models of Yang \etal \cite{YANG}, Tse \etal \cite{TSE}, Chen \etal \cite{CHEN} and ULP v0 \cite{ULPv0} have a one-term Prony-series. 

We also note that FE models are not always using experimental results directly. There are a number of models which have opted to use optimisation schemes based on running many simulations and picking parameters which best reproduce experimentally determined histories, such as from the data of Hardy \etal \cite{Hardy2001}. 
This approach is problematic because the parameter optimisation results now depend upon intrinsic properties of the model such as the geometry. This means that even while using the exact same validations, different models can yield substantially different predictions \cite{Ji2014}.

Another common practice in FE models (and also in various experimental papers) is that the hyperelastic and viscoelastic response are separately modelled. These effects can be either additively decomposed using the theory of linear viscoelasticity (section \ref{sec:LV}) or multiplicatively decomposed using the theory of quasi-linear viscoelasticity (section \ref{sec:QLV}). Some groups tend to merge different experimental data:  they source viscoelastic properties and hyperelastic properties from different experiments and combine them together.
However, one can obtain different fits for each hyperelastic model; this can be seen in the work of MacManus \etal \cite{MacManus2017b} and Eskandari \etal \cite{Eskandari2020}. 
Furthermore, using different hyperelastic models essentially amounts to altering the value of $M_0$. 
This is frequently done in FE models e.g. the KTH model \cite{Kleiven2007} simply scales the data by a factor of 2, and the ULP v1 \cite{ULPv1b} model scales the data from the results of the 1977 paper of Khalil \etal \cite{Khalil1977}. However, this scaling affects both the attenuation and  dispersion and thus changes the mechanical behaviour of the tissue compared to the original model from the original experimental paper.

\begin{figure}[h!]
    \centering
      
         \includegraphics[trim = 90 150 20 20, clip, width=0.95\textwidth]{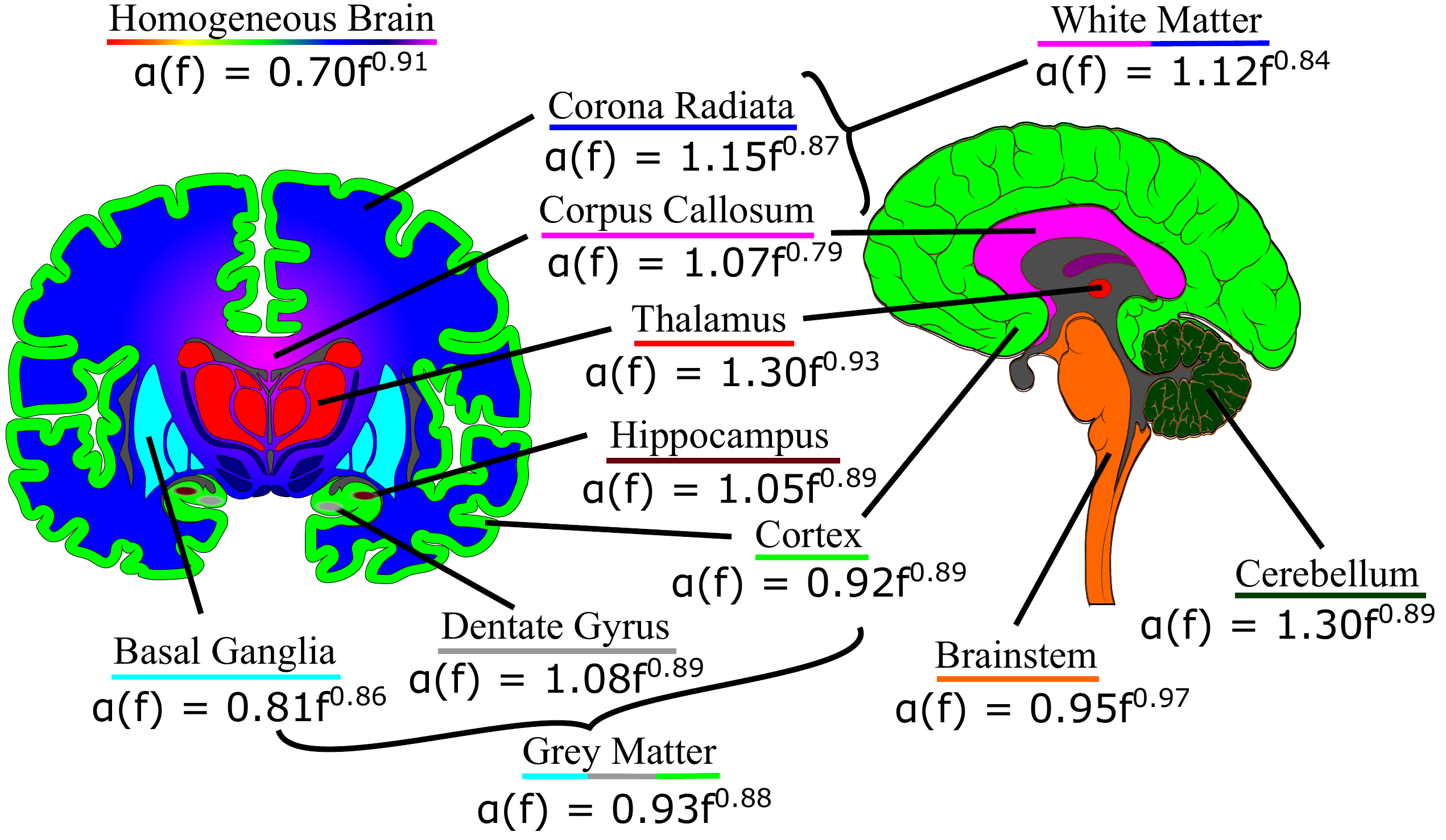}
          \caption{Schematic of brain sections coronal (left) and saggital (right) with twelve different attenuation power laws with nine different regions (basal ganglia, brainstem, corona radiata, corpus callosum, cortex, dentate gyrus, hippocampus, and thalamus)
          and the remaining three summarising the white matter, grey matter and the homogeneous brain. }         
          
    \label{fig:brain-images}
\end{figure}

\begin{table}[t]
\begin{center}
\begin{threeparttable}[t]
\small{
\begin{tabular}{ | p{4.3cm} | p{9mm} | p{1.5cm} | p{7.5cm} | }
 \hline
 \textbf{Reference} & \textbf{Year} & \textbf{Species} & \textbf{FE Model(s)}\\
 \hline \hline
MacManus \etal \cite{MacManus2017a} & 2017 & Rat & UCD v2 \cite{UCDv2}\\
\hline
Miller \etal \cite{ATLAS}$^b$ & 2016 & - & ATLAS \cite{ATLAS}\\
 \hline
 Rashid \etal \cite{Rashid2012b} & 2012 & Pig &  Khanuja-Unni \cite{KHANUJA}, YEAHM \cite{YEAHMa, YEAHMb}\\
\hline
 Kleiven \cite{Kleiven2007} (using data from Nicolle \etal \cite{Nicolle2005}) & 2005 & Pig &  ADAPT \cite{ADAPT}, ICM  \cite{ICM}, KTH v2 \cite{KTHv2}\\
 \hline
Cloots \etal \cite{Cloots2013} (using data from Nicolle \etal \cite{Nicolle2005}) & 2005 & Pig &  ANISO-KTH v1 \cite{ANISOKTHv1}, ANISO-KTH v2 \cite{ANISOKTHv2}, WHIM v1 \cite{WHIMv1}\\
\hline
 Zhang \etal \cite{Zhang2004}$^a$ & 2004 & - & Chen \& Ostoja-Starzewski \cite{CHEN}
 \\
 \hline
 Willinger \& Baumgartner \cite{ULPv1a}$^a$ & 2003 & - & ULP v1 \cite{ULPv1a, ULPv1b}
 \\
 \hline
Takhounts \etal \cite{Takhounts2003} & 2003 & Human & Cai \etal \cite{CAI}, SIMon v1 \cite{SIMONv1}  \\
 \hline
 Zhang \etal \cite{WSUBIM}$^b$ & 2001 & - & Tse \etal \cite{TSE}, UCD v2 \cite{UCDv2}, WSUBIM \cite{WSUBIM}, Yang \etal \cite{YANG} \\
 \hline
 Willinger \etal \cite{ULPv0} (using data from Shuck \& Advani \cite{Shuck1972}) & 1972 & Human & Tse \etal \cite{TSE}, ULP v0 \cite{ULPv0}, Yang \etal \cite{YANG}
 \\
 \hline
 Zhang \etal \cite{Zhang2001} (using data from Shuck \& Advani \cite{Shuck1972}) & 1972 & Human & Tse \etal \cite{TSE}, Yang \etal \cite{YANG}
 \\
 \hline
  Zhao \& Ji \cite{WHIMv2b} (using data from Shuck \& Advani \cite{Shuck1972}) & 1972 & Human & WHIM v2 \cite{WHIMv2a, WHIMv2b}
 \\
 \hline
  Mendis \etal \cite{Mendis1995} (using data from Estes \& McElhaney \cite{Estes1970})  & 1970 & Human & Subramaniam \etal \cite{SUBRAM}, UCD v1 \cite{UCDv1}\\
 \hline
\end{tabular}
\begin{tablenotes}
  \item[a] \emph{No experimental viscoelastic source was found}
  \item[b] \emph{These papers use optimised parameters selected to match experiment results and thus thus do not come directly from experimental viscoelastic data. For example, Zhang \etal \cite{Zhang2001} used datasets such as pressure data by Troseille \etal \cite{Trosseille1992} and Nahum \etal \cite{Nahum1977} for optimisation.}
\end{tablenotes}
\caption{Sources of experimental viscoelastic data (in chronological order) used by 19 current state-of-the-art FE models for the obtention of \emph{dimensionless} Prony-series $\hat{g}(t)$}
\label{tab:FEMdatasources}
}
\end{threeparttable}
\end{center}
\end{table}

We performed a thorough literature review of experimental papers and of computational simulation papers, with a total of more than 100 research articles. 
Most of the finite element method (FEM) based numerical solvers use viscoelastic material properties from the thirteen papers presented in  \autoref{tab:FEMdatasources}. 
Many of the FEM solvers currently assume that brain is a homogeneous material, and  only implement a single-term Prony-series, mostly with the assumption of linear viscoelasticity. Some recent FEM implementations use the QLV implementation. 

We gathered the viscoelastic properties, specifically, the Prony-series parameters implemented in the FEM solvers as well as those recorded in the experimental papers for different tissue types, namely: 1) homogeneous brain, 2) brainstem, 3) basal ganglia, 4) cerebellum, 5) corona radiata, 6) corpus callosum, 7) cortex, 8) dentate gyrus, 9) hippocampus, 10) thalamus, 11) grey matter and 12) white matter. These regions are depicted in \autoref{fig:brain-images}. A total of 8 different animals were considered: namely, pig, rat, human, mouse, cow, sheep, monkey and dog.
In the main article, we provide a detailed analysis of the viscoelastic behaviour of the homogeneous brain as used in FEM solvers, and relegate the viscoelastic properties of other tissue types to the supplementary material. 

Some models include anisotropy \cite{Ning2006, Eskandari2021, ANISOKTHv1, ANISOKTHv2, WHIMv2a, WHIMv2b} or porosity \cite{Hosseini-Farid2020, Forte2017} in addition to linear or quasi-linear viscoelasticity, but we did not report these effects (in general, fibre reinforcement does not contribute significantly to the mechanical response in the parallel or perpendicular shearing directions \cite{Ning2006}). 
Similarly, we ignored compressibility because brain matter is mostly incompressible \cite{Libertiaux2011}.

Furthermore, experimental papers oftentimes provided multiple Prony-series fits for the same region, but with differing strain rates \cite{Li2020, Qian2018, Ramzanpour2020}, strains \cite{Ramo2018a, Ramo2018b, Shetye2014}, indentation depths \cite{Shafieian2009}, loading rates \cite{Gefen2004}, impact angle \cite{Qiu2020}, velocity \cite{Qiu2020}, loading modes (e.g. tension, compression, shear etc.) \cite{Budday2017a}, direction relative to fibres \cite{Eskandari2021, Li2021}, loading cycle \cite{Budday2017b}, boundary condition \cite{Cheng2007}, preconditioning or no preconditioning \cite{Budday2017a, Gefen2004, Shafieian2009}, injured or uninjured tissue \cite{Shafieian2009}, plane of experiment \cite{MacManus2017a, Li2021, Sundaresh2021}, and animal age \cite{Elkin2011a, Elkin2013, Finan2012a, Finan2012b, MacManus2017a}. 
Having many Prony-series come from a given study was not desirable because that study would disproportionally affect the final, averaged, results and because it is well known from previous literature reviews that viscoelastic parameters may vary immensely from one study to another \cite{Budday2020, Hrapko2008, Chatelin2010}. 
As a result, we decided to take fits from mature and uninjured tissue only. 
When choice was available, we took fits for the highest strain rate, strain, indentation depth and velocity. 
When available, we took data for all modes, the first loading cycles, and no slip boundary conditions. 
Fits in directions orthogonal fibres were also preferred, to neglect anisotropic effects. 
Data from the axial plane was preferred because slices are more homogeneous along this plane; if that data was unavailable, then the sagittal plane was taken instead. Finally, if neither of these were available, the coronal plane was taken.
Impact angles of 0 degrees were also preferred. 
Lastly, preconditioned fits were taken when available, as they were observed to be closer to the other data, and also some unconditioned fits were found to have $M_\infty=0$ \cite{Shafieian2009}, which it unphysical as is corresponds to a fluid.
Cases where different Prony-series were provided for different locations within the same region were kept, and highest order Prony-series were taken in all cases. When differing fits were provided for different animals \cite{MacManus2017a, MacLean2010, Takhounts2003}, these were also kept and the animal type was recorded.
 
For Prony-series fits used in FE models, we found that only a single FE model provided specific viscoelastic parameters for the corpus callosum \cite{CAI}. This fit was thus considered as white matter owing to a lack of other data. Similarly, only one FE model provided viscoelastic properties for the cerebrum \cite{TSE}, so this was included within the homogeneous brain data for FE models.

Following the literature review, we focused on a total of six key quantities: the relaxation function $g(t)$, storage modulus $M^\prime(\omega)$, loss modulus $M^{\prime\prime}(\omega)$, inverse quality factor $Q^{-1}(\omega)$, dispersion $c(\omega)$ and attenuation $\alpha(\omega)$. 

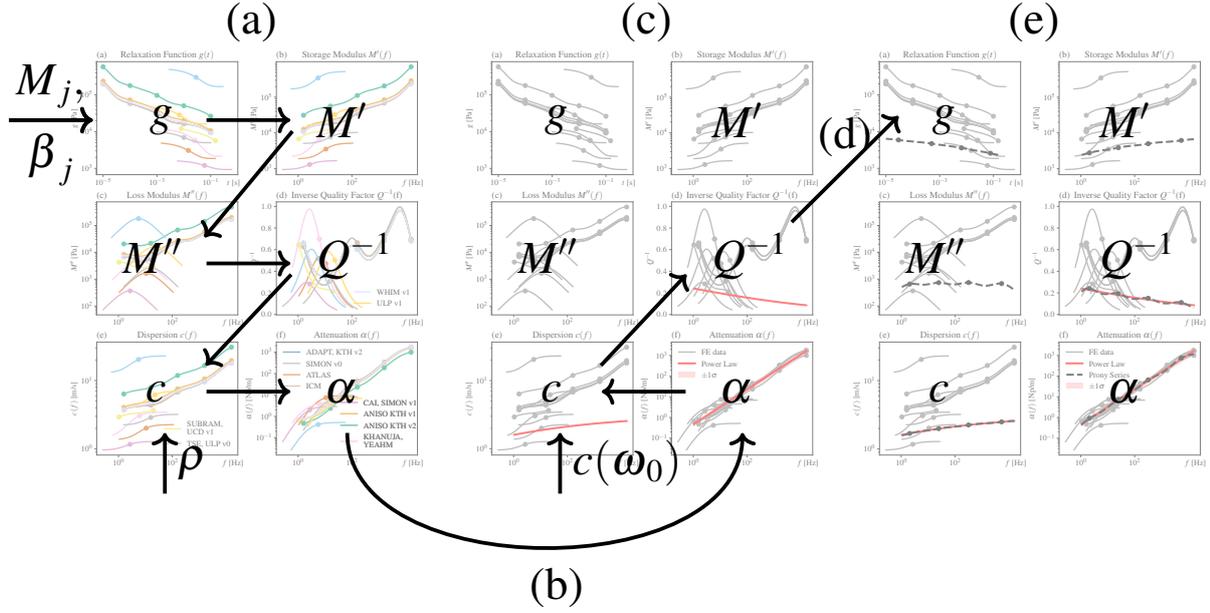
\begin{figure}[h!]
\begin{tikzpicture}
\begin{scope}[xshift=0cm]
    \node[anchor=south west,inner sep=0, opacity=0.5] (image) at (0,0) {
    \includegraphics[ width=0.28\textwidth]{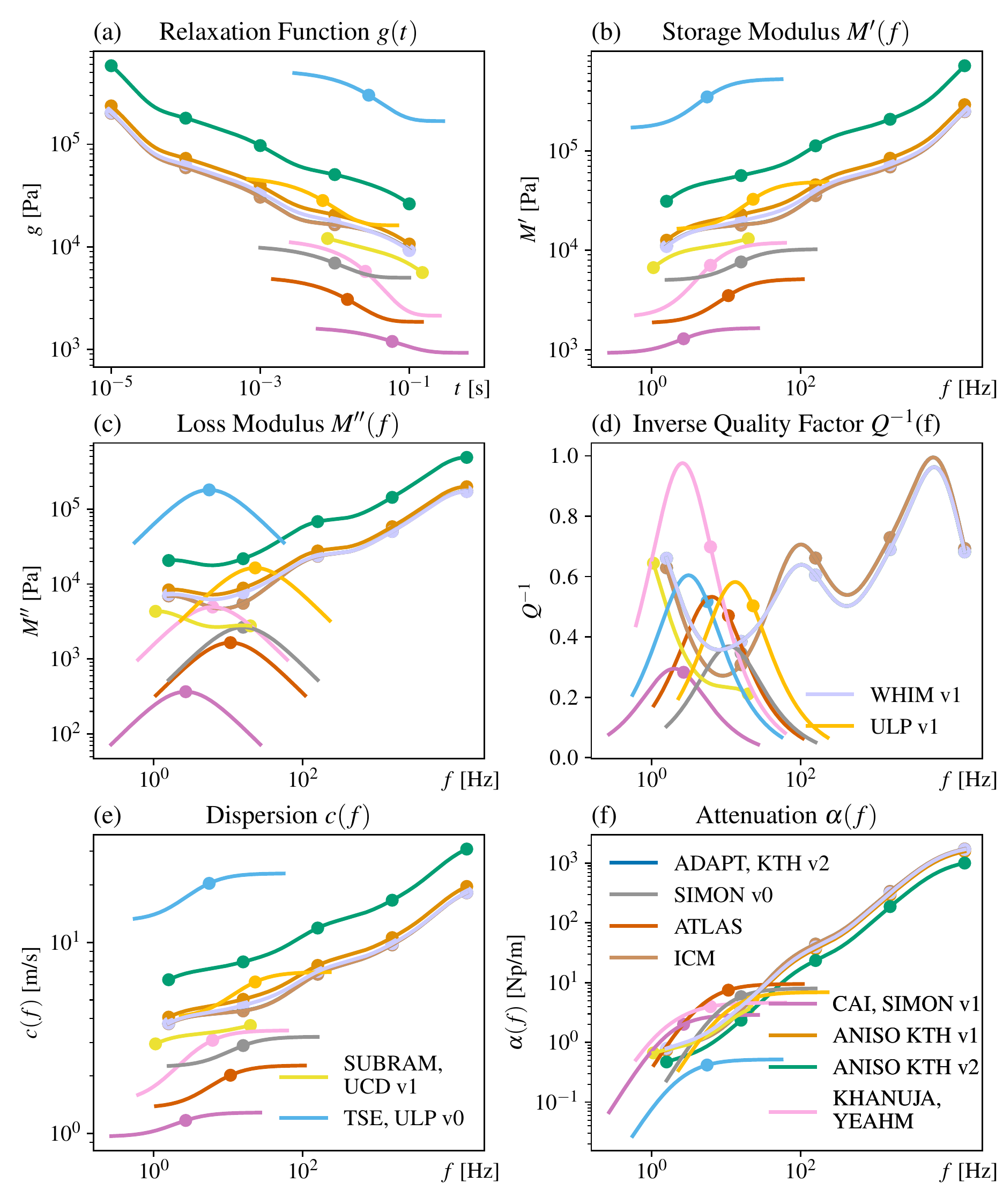}};
    \node[anchor=south west,inner sep=0, opacity=0.5] (image) at (5.2,0) {
    \includegraphics[ width=0.28\textwidth]{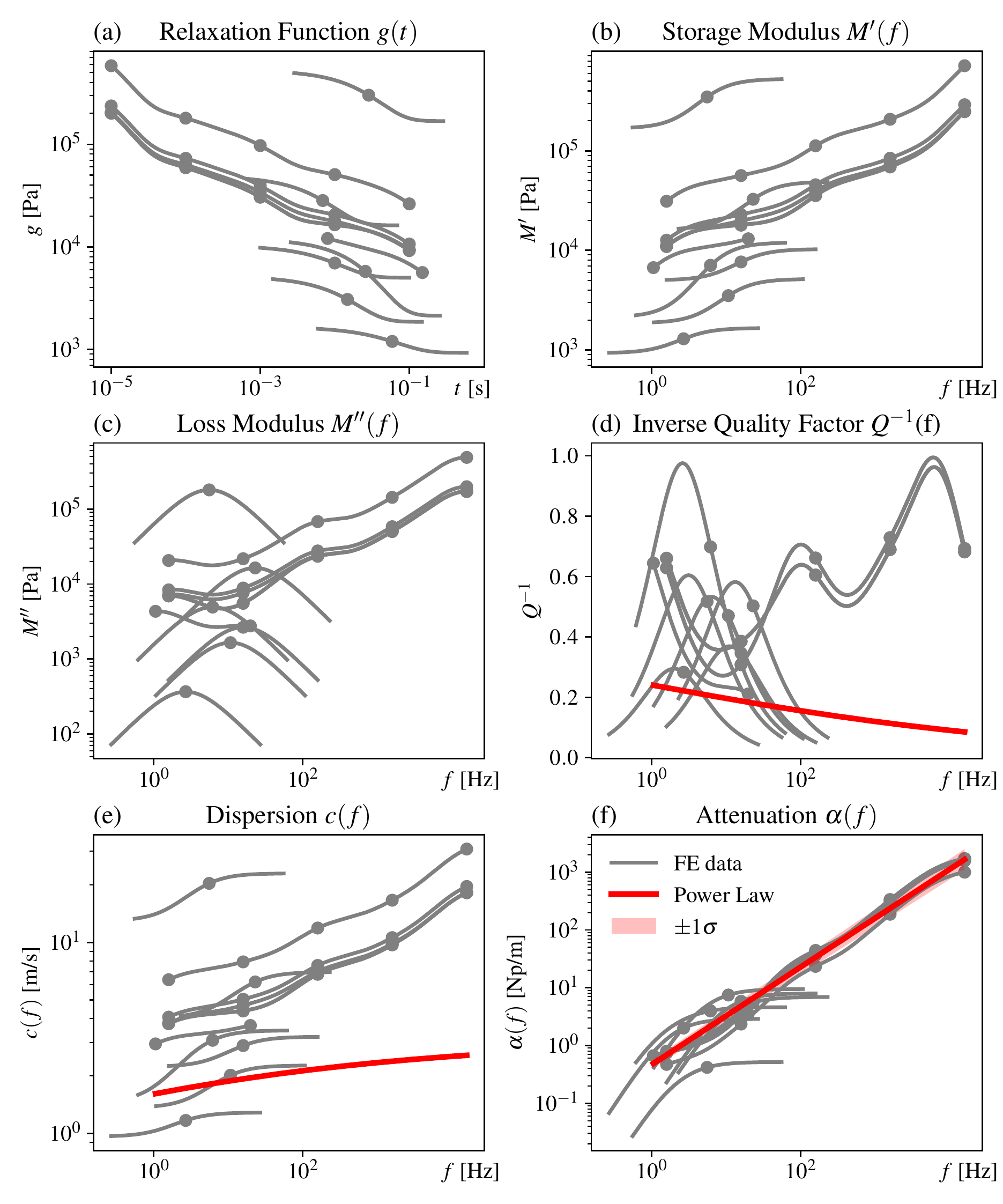}};
    \node[anchor=south west,inner sep=0, opacity=0.5] (image) at (10.3,0) {
    \includegraphics[ width=0.28\textwidth]{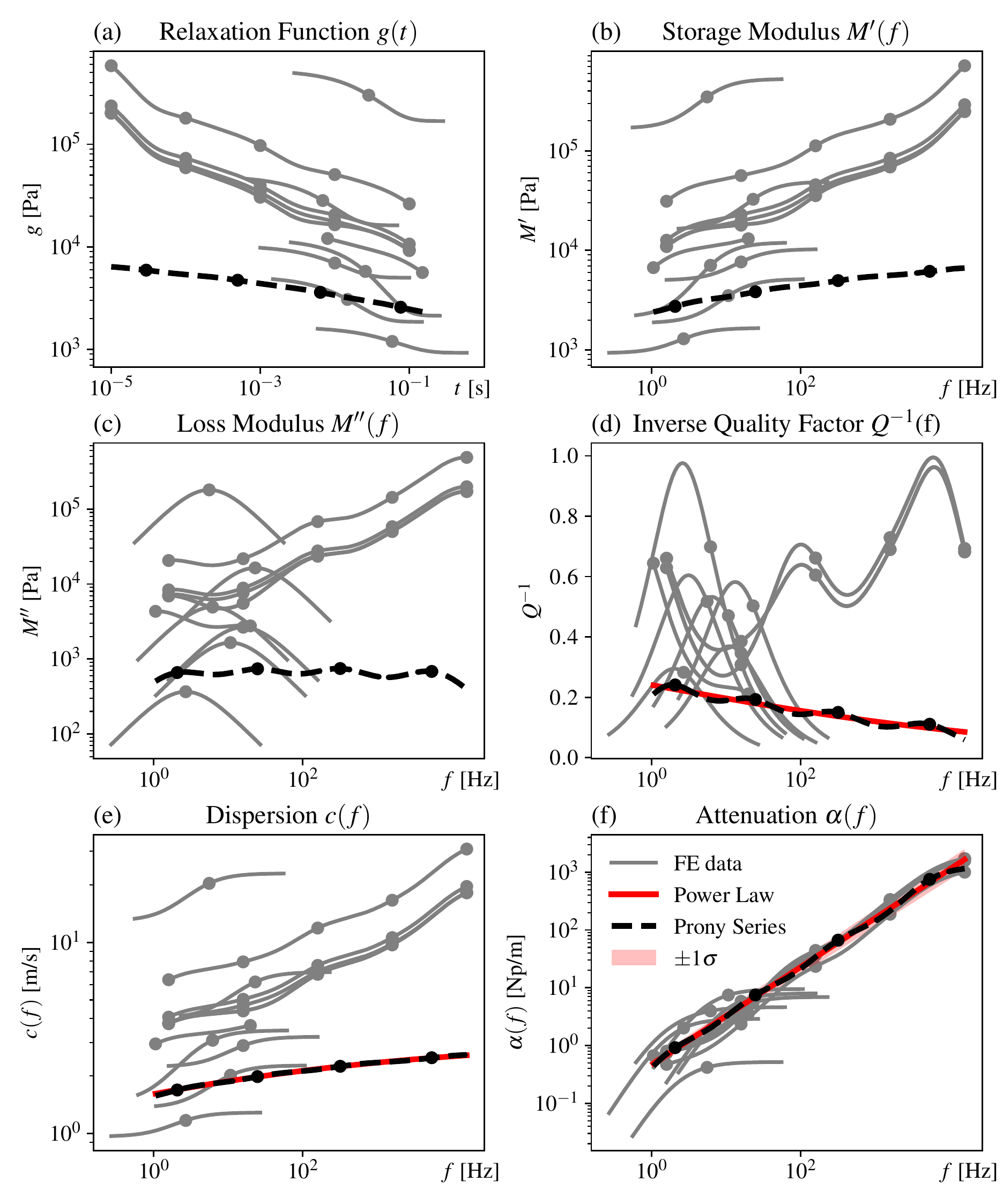}};
    
    \node [anchor=west] (in) at (2,6) {\Large (a)};
    \node [anchor=west] (in) at (7.2,6) {\Large (c)};
    \node [anchor=west] (in) at (12.3,6) {\Large (e)};
    \node [anchor=west] (in) at (6,-1.5) {\Large (b)};
    
    \node [anchor=west] (in) at (1,4.7) {\Large $\boldsymbol{g}$};
    \node [anchor=west] (in) at (0.65,2.9) {\Large $\boldsymbol{M''}$};
    \node [anchor=west] (in) at (0.95,1.1) {\Large $\boldsymbol{c}$};
    \node [anchor=west] (in) at (3.2,4.7) {\Large $\boldsymbol{M'}$};
    \node [anchor=west] (in) at (3.2,2.9) {\Large $\boldsymbol{Q^{-1}}$};
    \node [anchor=west] (in) at (3.3,1.1) {\Large $\boldsymbol{\alpha}$};
    
    \node [anchor=west] (in) at (1+5.2,4.7) {\Large $\boldsymbol{g}$};
    \node [anchor=west] (in) at (0.65+5.2,2.9) {\Large $\boldsymbol{M''}$};
    \node [anchor=west] (in) at (0.95+5.2,1.1) {\Large $\boldsymbol{c}$};
    \node [anchor=west] (in) at (3.2+5.2,4.7) {\Large $\boldsymbol{M'}$};
    \node [anchor=west] (in) at (3.2+5.2,2.9) {\Large $\boldsymbol{Q^{-1}}$};
    \node [anchor=west] (in) at (3.3+5.2,1.1) {\Large $\boldsymbol{\alpha}$};
    
    \node [anchor=west] (in) at (1+10.3,4.7) {\Large $\boldsymbol{g}$};
    \node [anchor=west] (in) at (0.65+10.3,2.9) {\Large $\boldsymbol{M''}$};
    \node [anchor=west] (in) at (0.95+10.3,1.1) {\Large $\boldsymbol{c}$};
    \node [anchor=west] (in) at (3.2+10.3,4.7) {\Large $\boldsymbol{M'}$};
    \node [anchor=west] (in) at (3.2+10.3,2.9) {\Large $\boldsymbol{Q^{-1}}$};
    \node [anchor=west] (in) at (3.3+10.3,1.1) {\Large $\boldsymbol{\alpha}$};
    
    \node [anchor=west] (in) at (-1,4.7) {};
    \node [anchor=west] (rho) at (1.2,-0.4) {};
    \node [anchor=west] (c_1_left) at (1.2,0.7) {};
    \node [anchor=west] (g_1_start) at (0.4,4.7) {};
    \node [anchor=west] (g_1_end) at (1.6,4.7) {};
    \node [anchor=west] (M_1_start) at (3,4.7) {};
    \node [anchor=west] (M_1_end) at (1.6,3.0) {};
    \node [anchor=west] (M_2_start) at (1.6,2.8) {};
    \node [anchor=west] (M_2_end) at (3,2.8) {};
    \node [anchor=west] (c_1_start) at (1.6,1.3) {};
    \node [anchor=west] (c_1_end) at (1.6,1.1) {};
    \node [anchor=west] (a_1_start) at (3,1.1) {};
    \node [anchor=west] (a_1_end) at (3.6,0.7) {};
    
    \node [anchor=west] (a_2_in) at (8.8,0.7) {};
    \node [anchor=west] (M_3_end) at (8.2,2.8) {};
    \node [anchor=west] (c_2_start) at (6.8,1.3) {};
    \node [anchor=west] (c_2_end) at (6.8,1.1) {};
    \node [anchor=west] (a_2_start) at (8.2,1.1) {};
    
    \node [anchor=west] (ref_c) at (6.4,0.7) {};
    \node [anchor=west] (c_in_2) at (6.4,-0.4) {};
    
    \node [anchor=west] (in_3) at (11,4.9) {};
    \node [anchor=west] (out_3) at (9.3,3.2) {};

    \node [anchor=west] (g_3_end) at (11.9,4.7) {};
    \node [anchor=west] (M_4_start) at (13.3,4.7) {};
    \node [anchor=west] (M_4_end) at (11.9,3.0) {};
    \node [anchor=west] (M_5_start) at (11.9,2.8) {};
    \node [anchor=west] (M_5_end) at (13.3,2.8) {};
    \node [anchor=west] (c_3_start) at (11.9,1.3) {};
    \node [anchor=west] (c_3_end) at (11.9,1.1) {};
    \node [anchor=west] (a_3_start) at (13.3,1.1) {};
    \node [anchor=west] (a_3_end) at (13.9,0.7) {};
    
    \node [anchor=west] (c_3_in) at (11.5,0.7) {};
    \node [anchor=west] (rho_c_3) at (11.5,-0.4) {};
    
    \draw[->, line width=.5mm] (in) -- (g_1_start) node[midway,fill=none, above] {\Large $\boldsymbol{M}_{\boldsymbol{j}},$};
    \draw[->, line width=.5mm] (in) -- (g_1_start) node[midway,fill=none, below] {\Large $\boldsymbol{\beta}_{\boldsymbol{j}}$};
    \draw[->, line width=.5mm] (rho) -- (c_1_left) node[midway,fill=none, right] {\Large $\boldsymbol{\rho}$};
    \draw[->, line width=.5mm] (g_1_end) -- (M_1_start) node[midway,fill=none, above] {};
    \draw[->, line width=.5mm] (M_1_start) -- (M_1_end) node[midway,fill=none, above] {};
    \draw[->, line width=.5mm] (M_2_start) -- (M_2_end) node[midway,fill=none, above] {};
    \draw[->, line width=.5mm] (M_2_end) -- (c_1_start) node[midway,fill=none, above] {};
    \draw[->, line width=.5mm] (c_1_end) -- (a_1_start) node[midway,fill=none, above] {};
    
    \draw[<-, line width=.5mm] (M_3_end) -- (c_2_start) node[midway,fill=none, above] {};
    \draw[<-, line width=.5mm] (c_2_end) -- (a_2_start) node[midway,fill=none, above] {};
    \draw[<-, line width=.5mm] (ref_c) -- (c_in_2) node[midway,fill=none, right] {\Large $\boldsymbol{c}(\boldsymbol{\omega}_{\boldsymbol{0}})$};
    
    \draw[<-, line width=.5mm] (in_3) -- (out_3) node[midway,fill=none, above] {\Large (d)};
    
    \draw [->, line width=.5mm] (a_1_end) to [out=-90,in=-90] (a_2_in);
  
\end{scope}
\end{tikzpicture}
\caption{The workflow process consists of five steps: (a) forward calculation, (b) average power law calculation, (c) backward calculation, (d) averaged Prony-series fitting and (e) re-calculation.}
\label{fig:workflow}
\end{figure}

We adopted a workflow consisting of five key steps to analyse the different viscoelastic parameters extracted from the literature, see summary in \autoref{fig:workflow}. 
The steps are as follows.

\subsection{Forward calculation}
\label{subsec:forwardprop}
\begin{itemize}
 \item 
 The coefficients of the Prony-series, $M_j$ and $\beta_j=1/\tau_j$, $j=1,\ldots,N$, are recorded for each study during the literature review. These values together can be used to create the relaxation function $g(t)$ using equation \eqref{eq:prony}, or alternatively to calculate the dimensionless parameters $\hat{M}_j$ along with the instantaneous shear modulus $M_0$ via equation \eqref{eq:dimensionless_prony}. 
 
 \item 
 The Prony-series data is then used to calculate the storage modulus $M'(\omega)$ via equation \eqref{eq:storage} and the loss modulus $M''(\omega)$ via equation \eqref{eq:loss}. 
 
 \item 
 With the help of the loss and storage moduli, the inverse quality factor $Q^{-1}(\omega)$ is calculated using equation \eqref{eq:quality_def} along with the dispersion relation $c(\omega)$ via equation \eqref{eq:c_omega_from_prony}. 
 A mass density $\rho=1,000$ kg/m$^3$ was used for all tissues.
 
 \item 
 Using the quality factor and the dispersion, the attenuation power law $\alpha(\omega)$ is calculated from equation \eqref{eq:quality_relation}. 
 \end{itemize}
 
 \subsection{Average power law calculation} 
 \label{subsec:avgpowerlaw}
 Now attenuations are calculated for each of the Prony-series.  It is then possible to synthesise an averaged attenuation power law from these calculated curves.  Specifically, we conduct a linear fit in the log-log space using the $\alpha(\omega)$ laws evaluated only at the frequencies corresponding to their Prony-series decay coefficients $\omega = \beta_j,~j= 1,\ldots,N$. The valid frequency range of a fit was then taken to be $[\min_j \beta_j, \max_j \beta_j]$. Fits were only undertaken if there were at least three datapoints.
 
 \subsection{Backward calculation} 
 \label{subsec:backprop}
 Following the average power law calculation, we obtain an averaged power law $\alpha(\omega) = a \omega^b$. 
 Using a reference value of  $c = 2.1$ m/s at a frequency of 75 Hz derived from experiments on homogeneous brain tissue \cite{Espindola2017a, Tripathi2021}, it is possible to calculate the dispersion from the Kramers-Kronig relation as defined in equation \eqref{eq:KKC}. 
 The quality factor can then subsequently be calculated using the derived attenuation and dispersion laws via equation \eqref{eq:quality_relation}.
 
 \subsection{Averaged Prony-series fitting}
 \label{subsec:avgprony}
 The backward calculation yields quality factors valid over a range of angular frequencies $[\omega_1, \omega_2]$. 
 From this data, it is then possible to directly perform a curve fitting exercise for the dimensionless Prony-series parameters as per equation \eqref{eq:curve_fitting}. 
 The order of the Prony-series is set by equation \eqref{eq:order_prony}. 
 Here, we evaluated the inverse quality at 1,000 equally spaced points on log-scale, which we call logarithmically spaced points in the valid interval $[\omega_1, \omega_2]$. 
 Curve fitting with scipy's curve\_fit function \cite{Scipy} is then undertaken with the parameters $M_j$  constrained on $[0, 1]$ and the parameters $\beta_j$ constrained on $[\omega_1, \omega_2]$. 
 Because it was known that all parameters must be positive to be physically meaningful \cite{Carcione2007}, a softplus transform was also used to ensure this requirement. 
 Once the dimensionless parameters were fitted, we calculated the instantaneous shear modulus $M_0$ using equation \eqref{eq:get_M_0}. 
 Again, we used the references values $\rho=1,000$ kg/m$^3$ and $c=2.1$ m/s at 75 Hz . 
 Then the fitted Prony-series is  entirely defined.
 Lastly, we also provide the attenuation power laws used, so that users can conduct their own Prony-series fits if desired.

\section{Results and discussion}

\subsection{Anomalies in Prony-series}

In some cases, Prony-series predictions were found not in line with the general data. Specifically, a number of Prony-series were found to predict an inverse quality factor greater than 1. This has problematic physical implications - a dissipation factor greater than 1 would correspond to the case where more energy is dissipated than the total energy of the wave \cite{Carcione2007}. As a result, corrections to these series were required in order to make them comparable to the general data. To this end, the dimensionless series was truncated by entirely removing the highest frequency term in the series. All other terms in the series were left unchanged. With this, the quality was then found to be strictly below 1 for the range of valid frequencies as desired. However, conducting such a truncation will necessarily alter either the value of $M_0$, or of $M_\infty$. A choice thus must be made on which quantity to keep constant. In this work, $M_0$ was kept constant since this is a more robust experimental quantity than $M_\infty$. That is, it is physically impossible to measure $M_\infty$ since this would require waiting for an infinite amount of time. Thus, instead in experiments a large time is used instead to find the value at $t \to \infty$. However, since this cut-off time is arbitrary, this means that values of $M_\infty$ can vary. Furthermore, we found that the keeping $M_0$ constant yielded results more in line with our general findings.

\begin{figure}[h!]
    \centering
    \includegraphics[width=\textwidth]{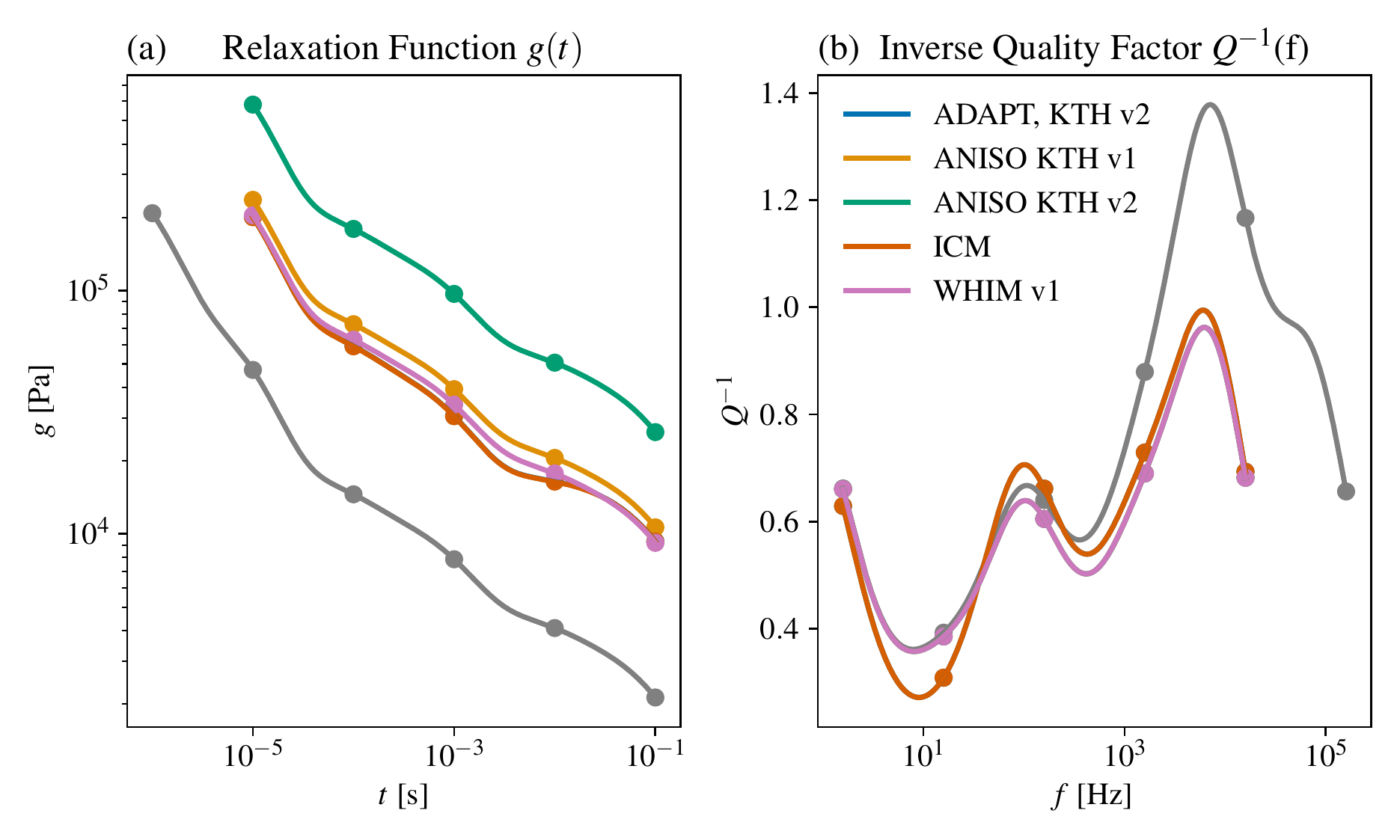}
    \caption{Anomalous Prony-series predictions (grey) and their corrected versions (coloured) {for (a) the relaxation function and (b) inverse quality factor.}}
    \label{fig:alter_Nicolle}
\end{figure}

Figure \ref{fig:alter_Nicolle} shows the original Prony-series (in grey color) and the truncated series (in color). These Prony-series are primarily derived from Nicolle's work \cite{Nicolle2004} and has been used by the ANISO KTH models and its variants, Imperial College, and Worchester models. As clearly evident from the right-subplot, the unphysical case of $Q^{-1}\geq1$ is present for high frequencies. On the other hand, the truncated Prony-series indeed produce $Q^{-1}<1$, however, this truncation does over estimate $g(t)$ with respect to the original series. Nevertheless, the truncated Prony-series produce results consistent with the average results. 

\subsection{Attenuation power laws in homogeneous brain}

\begin{figure}[H]
    \centering
    \includegraphics[trim= 10 10 0 10, clip, width=0.8\textwidth]{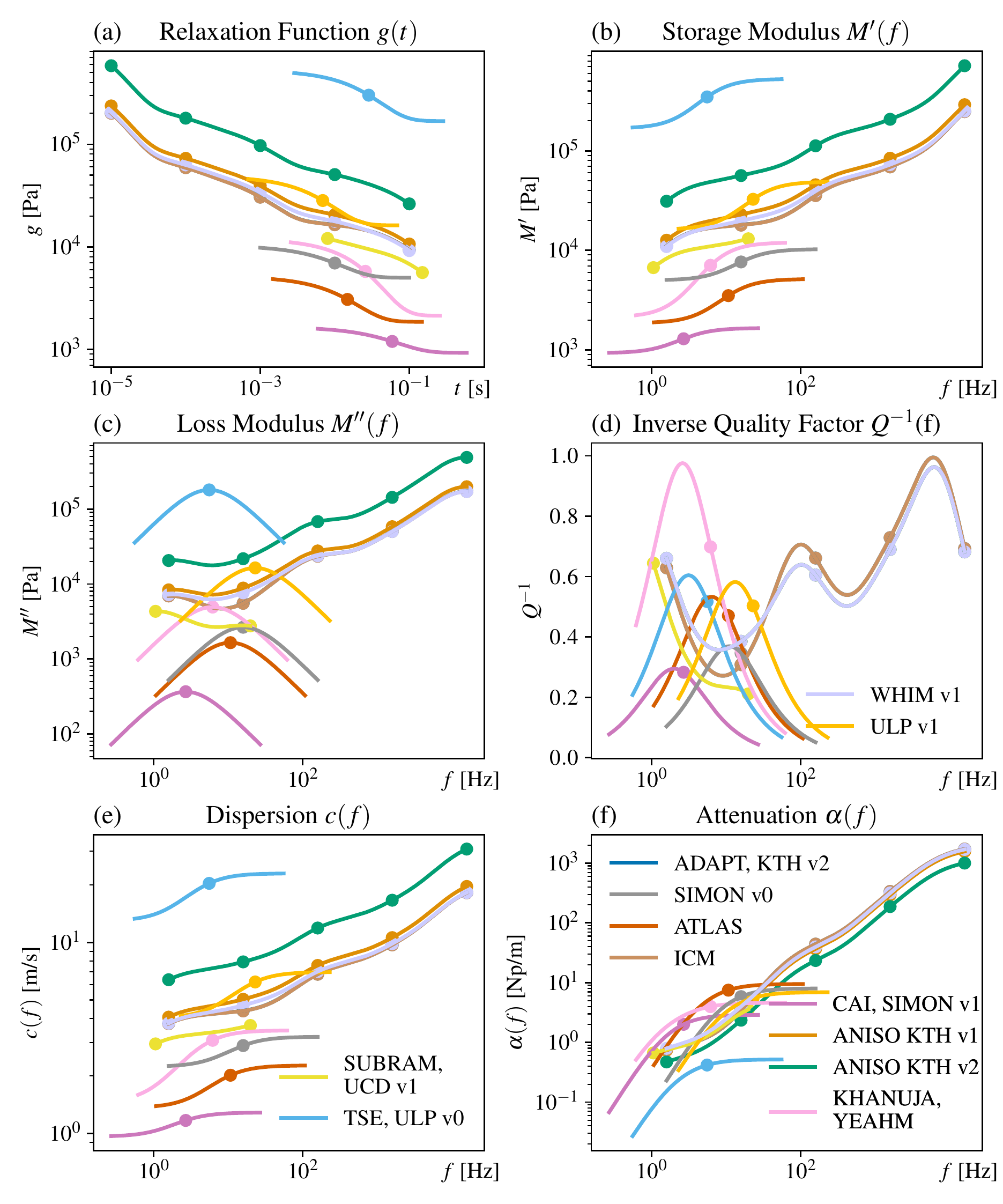}
    \caption{Derivation of attenuation and dispersion laws from each of the Prony-series used in FE models. {Shown are the predictions for (a) the relaxation function, (b) storage modulus, (c) loss modulus, (d) inverse quality factor, (e) dispersion and finally (f) attenuation.}}
    \label{fig:fem_brain_raw}
\end{figure}

Most of the computational models still use the homogeneous assumption while describing the viscoelastic properties of the brain matter. In this section, we consider the different Prony-series used in the common FE models describing the homogeneous brain deformation. 

The Prony-series collected were fitted using equation \eqref{eq:prony} which gives a continuous function as shown in \autoref{fig:fem_brain_raw}a. Most of the relaxation functions are close to each other except the one from Tse \etal \cite{TSE} and ULP v0 \cite{ULPv0} (light-blue), which uses the experimental data from Shuck and Advani \cite{Shuck1972}. 

We observed that most of the Prony-series are one-term, which greatly limits the frequency range they can capture. 
From the collected data, only models using data from Nicolle \etal \cite{Nicolle2005} are able to capture frequencies greater than 100 Hz. This limits the scope of possible applications. For example, road traffic and low-velocity missile impacts are associated with higher frequencies, on the order of 0.1-10 kHz \cite{Nicolle2004}. 

There is also a significant variation in the Prony-series data. Even FE models using the same experimental source can have different Prony-series fits. The variation is unsurprising given the experimental sources summarised in \autoref{tab:FEMdatasources}, many of which are 50 years old. This reliance on dated experimental data is problematic because experimental protocols have changed greatly over the past 50 years thanks to new experimental data and approaches \cite{Hrapko2008}. 
Studies indicating temperature and post-mortem time effects have lead to newer experimental approaches with better controls. 
For example, the data from the work of Shuck and Advani in 1972 \cite{Shuck1972} is an outlier, overestimating both the storage and loss moduli as compared to other studies \cite{Chatelin2010, Hrapko2008}. This data was obtained hours after autopsy, which itself may have been hours or days post-mortem. 
This issue is particularly problematic as it is well known that brain tissue stiffness increases quickly with post-mortem time. Weickenmeier \etal \cite{Weickenmeier2018} found that within 16 hours post-mortem, the loss and storage moduli were twice as stiff. 

Notwithstanding these extra considerations, there are large variations in the experimental protocol used in experiments in general \cite{Hrapko2008}, which makes it difficult to get consistency between results. 
However, as seen in \autoref{tab:FEMdatasources}, these older papers are some of the few experimental studies on human brain that are being used in FE models. 
The experimental data used by the UCD v2 \cite{UCDv2} model from MacManus \etal \cite{MacManus2017a} are obtained from experiments on rats, which is not ideal because the structure of the rodent brain is considerably different to that of a human \cite{Finan2012a}. Dai \etal \cite{Dai2018} recommend instead the use of experimental data from large animals (e.g. pig, rabbit, sheep, etc.) above rodents when data from human brains are not available, and Nicolle \etal \cite{Nicolle2004} report no significant difference in viscoelastic behaviour between porcine and human brain matter. 

Furthermore, the assumption that the brain is homogeneous with respect to viscoelastic properties is weak, as results can vary greatly depending on what region of the brain is being considered \cite{MacManus2017a}. It is thus important that the data for the homogeneous brain be taken from a representative region. However, the data of Nicolle \etal \cite{Nicolle2005} and of Shuck \& Advani \cite{Shuck1972} are in fact obtained from the corona radiata region. This is a white matter region which is mechanically quite different from the mixed white-grey matter region studied by Rashid \etal \cite{Rashid2012a}.

Predicted quantities from the collected Prony-series for the homogeneous brain as used in FE models are shown in \autoref{fig:fem_brain_raw}. In general, most Prony-series are only one-term series and span low frequency ranges, with the major exceptions of ADAPT \cite{ADAPT}, ANISO KTH v1 \cite{ANISOKTHv1}, ANISO KTH v2 \cite{ANISOKTHv2}, WHIM v1 \cite{WHIMv1}, ICM \cite{ICM} and KTH v2 \cite{KTHv2}, which use the data of Nicolle \etal \cite{Nicolle2005}. 
It is worth mentioning that the data from the YEAHM model \cite{YEAHMa, YEAHMb} and Khanuja \& Unni model \cite{KHANUJA} is a two-term Prony-series, coming from the fit of Rashid \etal \cite{Rashid2012b}. 
However, the decay coefficients $\beta_1 = 38.895 $ Hz and $\beta_2 = 38.911 $ Hz for this series are so close that an extended frequency range was also used to match a one-term Prony-series. 
Ignoring the extended range of the fit from Nicolle \etal \cite{Nicolle2005}, the data lies in the region of $t \in [10^{-3}, 10^{0}]$. 
Looking at \autoref{fig:fem_brain_raw}a, the data used by the Tse \etal and ULP v0 models \cite{TSE, ULPv0} 
and the models of Cai \etal and SIMon v1 \cite{CAI, SIMONv1} are the outliers. 
The data is found to span many logarithmic decades, and in fact shows a greater degree of variation in comparison to the review of Chatelin \etal \cite{Chatelin2010} which found data varying within almost two decades ($g(t) \in [20, 8000]$). Furthermore, the data from FE models is substantially stiffer than that of experimental papers, including both those of the review of Chatelin \etal, and also from this work (see supplementary materials). In contrast, the experimental data found in this work compares well to that in the review of Chatelin \etal, showing that is indeed an issue associated specifically with FE model data.

The outlier datasets of \autoref{fig:fem_brain_raw}a are worth further discussion. 
First, the major outlier is the series of Tse \etal \cite{TSE} and ULP v0 models \cite{ULPv0} (light blue), which comes from the study of Shuck \& Advani \cite{Shuck1972}.  As already discussed, this data is substantially stiffer than the rest of the literature.
Furthermore, there is a second outlier: the relaxation modulus data of Takhounts \etal \cite{Takhounts2003} used in the Cai \etal and SIMon v1 \cite{CAI, SIMONv1} models (purple) is lower than that in the rest of the literature. 
In that experiment, the tissue was stored by freezing and experimented on between 3 and 24 hours post-mortem. 
Other experimental sources used in FE models differ in this regard. For example, the experiment of Nicolle \etal \cite{Nicolle2005} (used in ADAPT \cite{ADAPT}, KTHv2 \cite{KTHv2}, ANISO KTH v1 \cite{ANISOKTHv1}, ANISO KTH v2 \cite{ANISOKTHv2}, WHIM v1 \cite{WHIMv1} and ICM \cite{ICM}) was conducted 24 hours post-mortem and that of Rashid \etal \cite{Rashid2012b} (used in the Khanuja-Unni \cite{KHANUJA} and YEAHM \cite{YEAHMa, YEAHMb} models) was conducted within 8 hours post-mortem. 
This may explain why the data of Cai \etal \cite{CAI} and SIMon v1 \cite{SIMONv1} is less stiff than the series of Nicolle \etal \cite{Nicolle2005}. 
The storage temperature is consistent with that of the other series of Rashid \etal \cite{Rashid2012b} and Nicolle \etal \cite{Nicolle2005}, which range from 4 to 6 C$^\circ$. 
This is important to note, because storage temperature can have a very large impact on the stiffness of brain tissue, with lower storage temperatures leading to stiffer behaviour \cite{Rashid2013b}. 
The temperature an experiment is conducted at is also important, with experiments conducted at room temperature showing a stiffer response than those measured at body temperature \cite{Hrapko2008}. Thus, it is worth noting that whilst the experiments of Rashid \etal \cite{Rashid2012b} and Takhounts \etal \cite{Takhounts2003} were conducted at room temperature, the work of Nicolle \etal \cite{Nicolle2005} was conducted at body temperature. Lastly, the specific region of the brain tested by Takhounts \etal \cite{Takhounts2003} is not listed, but we note that the study of Nicolle \etal \cite{Nicolle2005} was conducted on the corona radiata (white matter region) whilst that of Rashid \etal \cite{Rashid2012b} was conducted on mixed white and grey matter samples. This may in part explain why the data of Takhounts \etal \cite{Takhounts2003} appears to be an outlier.

The general disparity in the literature propagates through to the predictions of the storage and loss moduli where the same datasets are still outliers (\autoref{fig:fem_brain_raw}b and \autoref{fig:fem_brain_raw}c). In general, the storage modulus is observed to increase with frequency, as is the loss modulus. A characteristic n-shape is observed for the one-term Prony-series data predictions of the loss moduli, but this is simply due to the low order of the Prony-series and the use of an extended frequency range. For higher-term Prony-series such as that from Nicolle \etal \cite{Nicolle2005}, this behaviour is not observed. 

Importantly, a further conglomeration of the data is observed upon computation of the inverse quality (\autoref{fig:fem_brain_raw}d). 
This is a particularly important quantity to check as it is the ratio of the storage and loss moduli and is thus independent of the value of the instantaneous shear modulus $M_0$. 
The previous outlier datasets are found to lie within the rest of the data in terms of the inverse quality, which shows that the previous differences were largely due to their values for $M_0$. 
Oscillations in the inverse quality are also observed, which occurs due to a limited number of relaxation mechanisms in the Prony-series \cite{Emmerich1987, Blanc2016}. 
Approximately constant qualities are also anticipated due to the commonly used assumption of constant quality that is often made for determining regions of interest, as long as the dispersion is small \cite{Holm2019}.

Similarly, certain Prony-series predict very high wave speeds $c(\omega)$ of over 10 m/s (\autoref{fig:fem_brain_raw}e). 
Experiments have not observed speeds this high \cite{Jiang2015, Tripathi2021}. 
Instead, these predictions are caused by the high frequency behaviour of the Prony-series in the case of series derived from the data of Nicolle \etal \cite{Nicolle2005}, whilst for the data of Tse \etal and ULP v0 models \cite{TSE, ULPv0} it is caused by a very high instantaneous shear modulus.
{Despite all of this variation, the derived attenuation laws in \autoref{fig:fem_brain_raw}f are indeed generally observed to follow the expected power-law attenuation behaviour.}

\subsection{Attenuation power laws in heterogeneous brain}

\begin{figure}[h!]

    \centering
    \includegraphics[trim=10 10 1 10, clip, width=0.8\textwidth]{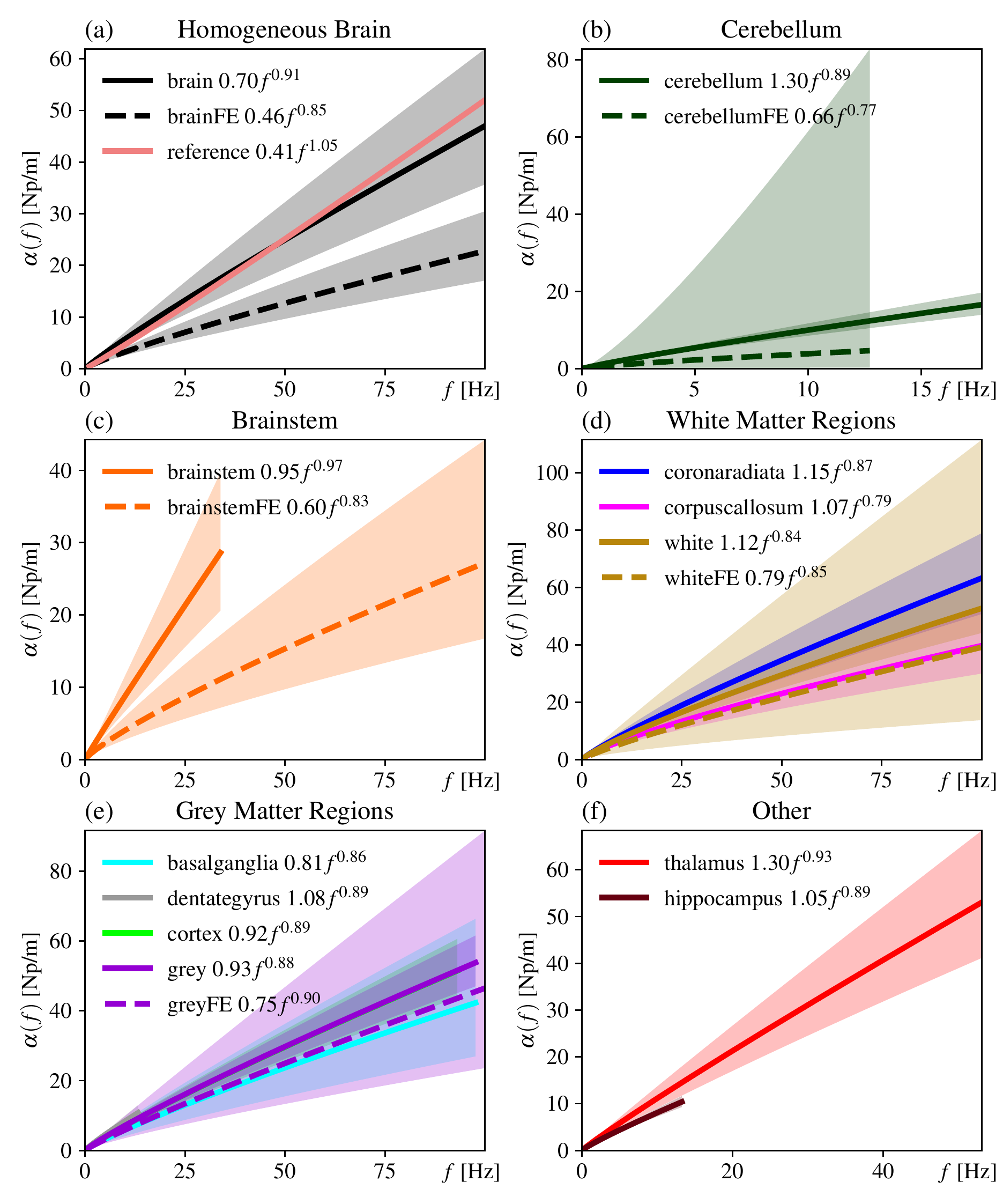}
    \caption{Averaged attenuation power laws at low frequencies ($<100$ Hz) for twelve different regions of the brain. Dashed lines refer to fits obtained from data used in FE models. Fits are only plotted over their respective valid frequency ranges. Shaded regions show one standard deviation $\pm 1\sigma$. {Plotted are (a) the homogeneous brain, (b) cerebellum, (c) brainstem, (d) white matter, (e) grey matter and (f) thalamus and hippocampus regions.}}
    \label{fig:low_freq_atten}
\end{figure}

We computed averaged attenuation power laws for twelve different regions in the brain, for  Prony-series from FE models and from recent experimental papers, following the same process shown for the homogeneous brain data of FE models. This yields averaged attenuation power laws and frequency intervals over which the fit is valid.
Both FE model data and experimental paper data were not always available for the all regions, but it was nonetheless possible to compare a number of key regions, as depicted in \autoref{fig:low_freq_atten}. Detailed calculations for each region are provided in the supplementary materials. Also, note that the the frequency axis is not the same for the tissue types as it is dependent on the experimental data. 

\autoref{fig:low_freq_atten}a shows the average attenuation law in the homogeneous brain tissue in FE models and the experiments together with a ``reference'' power law for homogeneous brain tissue we have used in our nonlinear (shock) shear wave modelling \cite{Tripathi2021}. The ``reference'' was obtained using the ultrasound shear wave imaging experiments performed on ex vivo porcine brain tissues \cite{Espindola2017a}. {The homogeneous brain tissue assumption is the most commonly used in FE modelling and so we get most of the data for the FE implementations from this region (12 unique Prony-series). For experimental data, 18 unique Prony-series were sourced for the homogeneous brain and this was not the most common tissue.} As evident the ``brain-FE'' law is significantly lower than the ``brain'' synthesised from the experimental data only. In fact, {a relative error calculation between the power-law attenuation of experimental data and FE models, calculated as $\displaystyle \frac{\left( a_{\rm Exp} \omega^{b_{\rm Exp}} - a_{\rm FE} \omega^{b_{\rm FE}}\right)}{a_{\rm Exp} \omega^{b_{\rm Exp}}}\times 100$, gives a range of errors of 43-52\% between 10 and 100 Hz, respectively. This suggests the need to revisit the viscoelastic modelling of brain matter in FE models to accurately capture the recent experimental data. 
The underestimate of attenuation in the FE implementation in contrast to experimental data is consistent in all the tissue types and the animal types. The lower attenuation in FE models tends to predict higher stiffness in contrast to the experimental data.

Surprisingly, the experimental power law closely aligns with the power-law attenuation we have used in our simulation studies \cite{Tripathi2021}. This could possibly be due to increased emphasis on high strain rate experiment in recent publications. 

Furthermore, there is also a greater degree of variation for FE model data, indicated by the larger ranges of uncertainty. The largest degree of uncertainty was found for the cerebellum region (\autoref{fig:low_freq_atten}b). This is not unexpected due to the lack of viscoelastic data in the FE model literature for the cerebellum region (only 4 unique Prony-series).

{For all regions except the brainstem and homogeneous brain (\autoref{fig:low_freq_atten}c), we observed that the fits from the experimental data lie within the error interval for the associated FE data. There are a number of possible reasons that may explain this. For one, as summarised in \autoref{tab:FEMdatasources}, the data used in FE models for the homogeneous brain largely comes from older viscoelastic sources, which give stiffer material properties as compared to recent experimental sources. Thus, for the homogeneous brain, it is unsurprising that FE models have significantly less attenuating power laws compared to recent experimental data. The question therefore becomes why we do not observe significant differences for the other regions. The reasons for this may be that the data for FE models is newer for these regions since heterogeneity is only implemented in recent FE models. Furthermore,  there is not much data for these regions, which leads to larger error intervals and thus less significant results -- specifically, from FE models there are only 6 unique Prony-series for white matter, 5 unique Prony-series for grey matter and 4 unique Prony-series for the cerebellum. By contrast, for the homogeneous brain, there are 12 unique Prony-series.}

For the white matter, in particular, we point out that the power laws in \autoref{fig:low_freq_atten}d for the corona radiata and corpus callosum are distinct (do not lie within the error regions of one another), but both of them lie within the error region for the white matter as used in FE models. {This underlines the importance of considering heterogeneity in FE models, instead of just white matter as a whole.} For comparison purposes, the corpus callosum data and corona radiata data were also pooled to create a single white matter region from experimental data, and this was found to also agree with the white matter data used in FE models. 

Similarly for grey matter in \autoref{fig:low_freq_atten}e, we found that the subregions of the basal ganglia, dentate gyrus and cortex all agreed with the grey matter data used in FE models. 
Moreover, the pooled data of the basal ganglia, dentate gyrus and cortex was used to generate a single grey matter region from experimental data and this was also found to agree with the data used for grey matter in FE models.

For two regions, namely the thalamus and the hippocampus (\autoref{fig:low_freq_atten}f), no reasonable comparison was possible with other FE model data since, to the best of our knowledge, these regions have not been modelled as viscoelastic materials in the FE models considered in this work. However, it is apparent that the thalamus is found to be the most attenuating region here and thus is mechanically different from other regions. This suggests that the thalamus is in fact an important region to include in FE models, and should not be neglected.

\subsubsection{Homogenisation of attenuation in brain}

\begin{figure}[h!]
    \centering
    \includegraphics[width=0.9\textwidth]{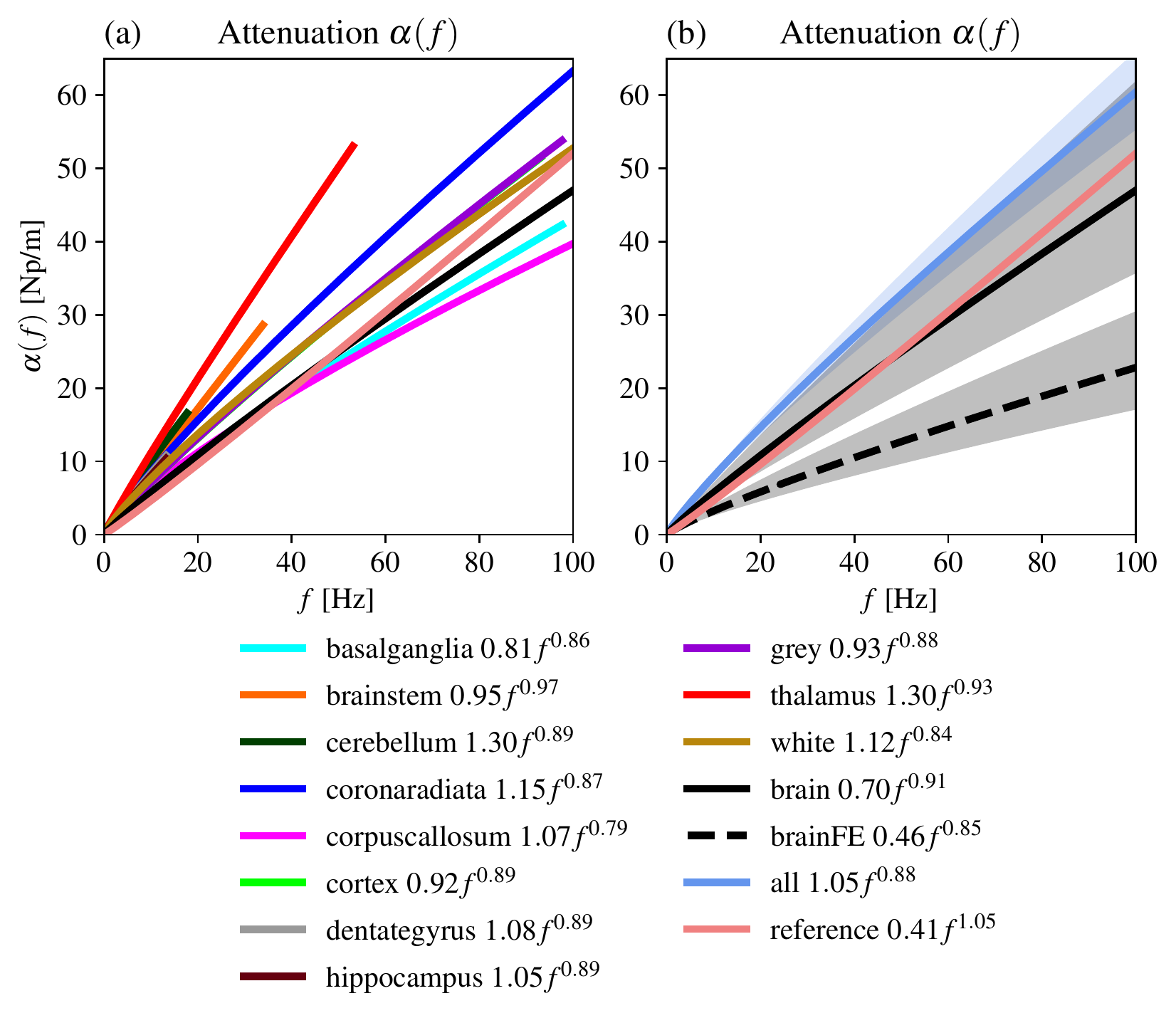}
    \caption{Attenuation power laws at low frequencies ($<100$ Hz) for experimental data. The ``all" fit is generated from merging the data for all regions bar the homogeneous brain (i.e. hippocampus, thalamus, brainstem, cerebellum, grey matter and white matter). Shaded regions show one standard deviation $\pm 1\sigma$. {Depicted are: (a) the power-law attenuation fits from experimental data for all of the gathered regions and (b) the power-law attenuation fits for the homogeneous brain (both FE and experimental), and the ``all'' region (solely experimental). The reference law of Tripathi \etal \cite{Tripathi2021} is also shown for comparison in (b).}}
    \label{fig:hetero}
\end{figure}
It is also important to test the validity of homogeneous brain measurements, because the brain is a highly heterogeneous tissue \cite{Antonovaite2021, Budday2020, Elkin2011a, Elkin2011b, Elkin2013, Finan2012a, Finan2012b, MacManus2017a, MacManus2017b}. In fact, one the key challenges identified in current FE modelling is the obtention of accurate heterogeneous data for models \cite{MacManus2022, Griffths2022}.
To this end, we pooled (referred as ``all'') the experimental data for all regions except the homogeneous brain to reconstruct the power law for the homogeneous brain from heterogeneous brain data. This was used to quantify the  the variation in the power law resulting with the assumption of homogeneous brain and the one constructed using the heterogeneous data. The results of these processes are shown in \autoref{fig:hetero}.

As expected, different tissue types in brain have different power laws  and the homogeneous brain power law (black curve) lies in between the different laws as seen in \autoref{fig:hetero}a. Note these laws are generated using the experimental data (and has no contribution from data collected from FE models). 
Also interesting to note is that the power law description for white and grey matter are almost overlapping as evident from \autoref{fig:hetero}a (see \autoref{tab:powerlaws_experi} for exact expressions). However, the Prony-series representation (for example: \cite{Hiscox2020}, \cite{Budday2015}) of these two regions are not as similar as their power laws. The reason for this overlap could be due to our averaging procedure over different experimental procedures, tissue types, temperature, animals, etc. are used in studies on white matter versus those on grey matter. Nevertheless, such an averaging is required in order to compare and leverage different experiments and to have some starting point for modelling nonlinear shear waves in brain. 
On the other hand, there have been discussions around the variations in elastic and anisotropic properties of white and grey matter.  
Many studies report conflicting results on the anisotropy of white matter (further discussion can be found in Budday \etal \cite{Budday2020}) and on  which tissue is stiffer (discussed in Zhang \etal \cite{Zhang2016}). 

Furthermore, we found that the ``all''  data (light blue) does not match the homogeneous brain data (solid black) as shown in \autoref{fig:hetero}b as good as the ``reference'' (light red), however, they are all still within each others' $\pm\sigma$. 
This difference could be due to the sampled regions for the homogeneous brain fits versus those of the rest of the experimental data. 
For example, the most common region found in our literature review is the cortex (43 unique Prony-series). However, since the locations for the homogeneous brain data are not explicitly given, it was not possible to determine whether the homogeneous brain data is dominated by the cortex data. However, this result nonetheless highlights a current discrepancy in the literature. It furthermore emphasises the need for considering the heterogeneity of the brain as opposed to attempting to construct a suitable averaged region, which can be highly subjective due to different averaging techniques.
However, these three curves: ``reference'', ``all'', or the homogeneous ``brain'' fit from experimental data are not within the error region of the  the attenuation power law from FE models (dashed black). 
Moreover, a relative error of $\displaystyle \frac{\left( a_{\rm Exp}\omega^{b_{\rm Exp}} - a_{\rm FE}\omega^{b_{\rm FE}}\right)}{a_{\rm Exp}\omega^{b_{\rm Exp}}}\times 100$ of 43-52\% between the power-law attenuation of experimental data and FE models from 10 to 100 Hz. Similarly, a relative error $\displaystyle \frac{\left( a_{\rm Hom} \omega^{b_{\rm Hom}} - a_{\rm All} \omega^{b_{\rm All}}\right)}{a_{\rm Hom} \omega^{b_{\rm Hom}}}\times 100$ of 29-39\% between the homogeneous brain and the ``all'' region from 10 to 100 Hz also indicates a discrepancy between the heterogeneous and homogeneous treatments of brain tissue. As before, it highlights the 
need to revisit the viscoelastic modelling of brain matter in FE models
to accurately capture the recent experimental data.

For completeness, the data for the fits for the experimental data and from FE model data are given for each region, and are shown in \autoref{tab:powerlaws_experi} and \autoref{tab:powerlaws_fem}, respectively. We have provided further details such as the raw data, full calculations and merged fits for each region in the supplementary materials.

\begin{remark}
Different Prony-series $G(t)$ are possible depending on the curve fitting procedure used. As a result we recommend users undertake their own curve fitting, but nonetheless we do provide our results in \autoref{tab:powerlaws_experi}, \autoref{tab:powerlaws_fem} and in the supplementary materials.
\end{remark}

\begin{table}[!h]
\begin{center}
\begin{threeparttable}[t]
\small{
\begin{tabular}{ | p{1.5cm} | p{1.9cm} | p{3.0cm} | p{5cm} | p{1.2cm} | }
 \hline
 \textbf{Region (\#Prony-series)} & \textbf{Angular Frequency Range [Hz]} & \textbf{Power Law $\ln \alpha = \ln{a} + b \ln \omega$ [Np/m]} & \textbf{Averaged Prony-series $\hat{g}(t)$} & \textbf{$\bfmM_0(\omega_0)$ [Pa]}\\
 \hline \hline
Brain (12) & [6.7, 1e+05] & $-2.32\pm0.13 + 0.85\pm0.02 \ln \omega$ & $0.32 + 0.165 e^{-13t} + 0.163 e^{-156t} + 0.172 e^{-2.01e+03t} + 0.18 e^{-3.39e+04t}$ & 6769 \\ \hline
Brain-stem (4) & [3.4, 1e+05] & $-2.03\pm0.20 + 0.83\pm0.04 \ln \omega$ & $0.227 + 0.164 e^{-7.59t} + 0.184 e^{-96.4t} + 0.207 e^{-1.38e+03t} + 0.217 e^{-2.72e+04t}$ & 7300 \\ \hline
Cere-bellum (4) & [3.3, 80] & $-1.82\pm0.82 + 0.77\pm0.47 \ln\omega$ & $0.475 + 0.525 e^{-15.2t}$ & 4411 \\ \hline
Grey (5) & [3.3, 7e+02] & $-1.94\pm0.21 + 0.90\pm0.07 \ln\omega$ & $0.253 + 0.241 e^{-13.4t} + 0.506 e^{-338t}$ & 4999 \\ \hline
White (6) & [13, 7e+02] & $-1.80\pm0.33 + 0.85\pm0.11 \ln\omega$ & $0.443 + 0.557 e^{-111t}$ & 4488 \\ \hline
\end{tabular}

\caption{Derived averaged attenuation power laws and corresponding Prony-series from averaged FE model data. {\#Prony-series refers to the number of \emph{unique} Prony-series. $M_0(\omega_0)$ is the instantaneous shear modulus obtained from a reference frequency of 75 Hz i.e. $\omega_0=150\pi$ Hz.}}
\label{tab:powerlaws_fem}
}
\end{threeparttable}
\end{center}
\end{table}

\begin{table}
\begin{center}
\begin{threeparttable}[t]
\small{
\resizebox{.8\textwidth}{!}{
\begin{tabular}{ | p{1.5cm} | p{1.9cm} | p{3.0cm} | p{5cm} | p{1.2cm} | }
 \hline
 \textbf{Region (\#Prony-series)} & \textbf{Angular Frequency Range [Hz]} & \textbf{Power Law $\ln \alpha = \ln{a} + b \ln \omega$ [Np/m]} & \textbf{Averaged Prony-series $\hat{g}(t)$} & \textbf{$\bfmM_0(\omega_0)$ [Pa]}\\
 \hline \hline
Basal ganglia (6) & [2e-05, 6.1e+02] & $-1.80\pm0.21 + 0.86\pm0.04 \ln \omega$ & $0.0118 + 0.0129 e^{-5.56e-05t} + 0.0225 e^{-0.000726t} + 0.0418 e^{-0.00917t} + 0.0765 e^{-0.118t} + 0.136 e^{-1.56t} + 0.233 e^{-21.6t} + 0.465 e^{-350t}$ & 4997 \\ \hline
Brain (18) & [0.01, 7.3e+02] & $-2.03\pm0.09 + 0.91\pm0.03 \ln \omega$ & $0.0872 + 0.0722 e^{-0.0385t} + 0.122 e^{-0.844t} + 0.22 e^{-17.6t} + 0.498 e^{-396t}$ & 5167 \\ \hline
Brain-stem (23) & [0.0048, 2.1e+02] & $-1.83\pm0.11 + 0.97\pm0.04 \ln\omega$ & $0.0752 + 0.0495 e^{-0.0204t} + 0.0868 e^{-0.479t} + 0.173 e^{-9.31t} + 0.616 e^{-195t}$ & 4642 \\ \hline
Cere-bellum (18) & [0.0061, 1.1e+02] & $-1.36\pm0.06 + 0.89\pm0.02 \ln(\omega)$ & $0.0408 + 0.0462 e^{-0.0207t} + 0.0898 e^{-0.339t} + 0.192 e^{-5.19t} + 0.632 e^{-92.9t}$ & 4464 \\ \hline
Dentate gyrus (7) & [0.079, 83] & $-1.55\pm0.06 + 0.89\pm0.03 \ln\omega$ & $0.107 + 0.111 e^{-0.248t} + 0.205 e^{-3.76t} + 0.577 e^{-63.6t}$ & 4435 \\ \hline
Cortex (43) & [2e-05, 5.8e+02] & $-1.71\pm0.05 + 0.89\pm0.02 \ln\omega$ & $0.0121 + 0.0114 e^{-5.75e-05t} + 0.019 e^{-0.00082t} + 0.0346 e^{-0.0111t} + 0.063 e^{-0.151t} + 0.115 e^{-2.03t} + 0.212 e^{-27.3t} + 0.532 e^{-427t}$ & 5290 \\ \hline
Corpus callosum (17) & [0.00099, 6.6e+02] & $-1.38\pm0.10 + 0.79\pm0.03 \ln \omega$ & $0.0103 + 0.0217 e^{-0.00345t} + 0.0541 e^{-0.0476t} + 0.131 e^{-0.676t} + 0.275 e^{-10.7t} + 0.508 e^{-221t}$ & 4682 \\ \hline
Corona radiata (19) & [0.0016, 1e+05] & $-1.46\pm0.09 + 0.87\pm0.02 \ln \omega$ & $0.00459 + 0.00549 e^{-0.00527t} + 0.0106 e^{-0.0842t} + 0.0219 e^{-1.29t} + 0.0454 e^{-19.7t} + 0.0951 e^{-297t} + 0.202 e^{-4.42e+03t} + 0.615 e^{-7.73e+04t}$ & 24390 \\ \hline
Hippo-campus (18) & [0.01, 83] & $-1.58\pm0.04 + 0.89\pm0.02 \ln \omega $ & $0.0798 + 0.101 e^{-0.0517t} + 0.216 e^{-1.45t} + 0.604 e^{-42.9t}$ & 4421 \\ \hline
Grey (56) & [2e-05, 6.1e+02] & $-1.69\pm0.05 + 0.88\pm0.01 \ln \omega$ & $0.0114 + 0.011 e^{-5.79e-05t} + 0.0186 e^{-0.000828t} + 0.034 e^{-0.0113t} + 0.0625 e^{-0.154t} + 0.115 e^{-2.09t} + 0.213 e^{-28.2t} + 0.534 e^{-445t}$ & 5349 \\ \hline
Thalamus (11) & [0.0047, 3.3e+02] & $-1.46\pm0.08 + 0.93\pm0.03 \ln \omega$ & $0.0349 + 0.0334 e^{-0.0231t} + 0.0681 e^{-0.609t} + 0.157 e^{-13.5t} + 0.707 e^{-333t}$ & 5052 \\ \hline
White (36) & [0.00099, 1e+05] & $-1.43\pm0.07 + 0.84\pm0.02 \ln \omega$ & $0.00528 + 0.0068 e^{-0.00245t} + 0.0124 e^{-0.0258t} + 0.0243 e^{-0.261t} + 0.0466 e^{-2.69t} + 0.0857 e^{-28.6t} + 0.149 e^{-318t} + 0.242 e^{-3.78e+03t} + 0.427 e^{-5.46e+04t}$ & 13909 \\ \hline
\end{tabular}%
}

\caption{Derived averaged attenuation power laws and corresponding Prony-series from averaged experimental data. {\#Prony-series refers to the number of \emph{unique} Prony-series. $M_0(\omega_0)$ is the instantaneous shear modulus obtained from a reference frequency of 75 Hz i.e. $\omega_0=150\pi$ Hz. The white region is formed from merging the corona radiata and corpus callosum regions, and the grey region is formed from merging the cortex, dentate gyrus and basal ganglia regions.}}
\label{tab:powerlaws_experi}
}
\end{threeparttable}
\end{center}
\end{table}
\subsection{Variation in attenuation due to animal selection}

There are multiple factors that can cause differences in experimental results, but one that is of key importance is that of the suitability of surrogate animals. 
The influence of this factor is vital to check because fresh human brain tissue is far more difficult to source than tissue from other animals such as pigs or cows. 
Thus, it is necessary to ascertain if surrogate tissues can be used {since this will have major ramifications on the ease of obtention of suitable experimental data.}

\begin{figure}[h!]
    \centering
    \includegraphics[trim=10 10 1 10, clip, width=0.8\textwidth]{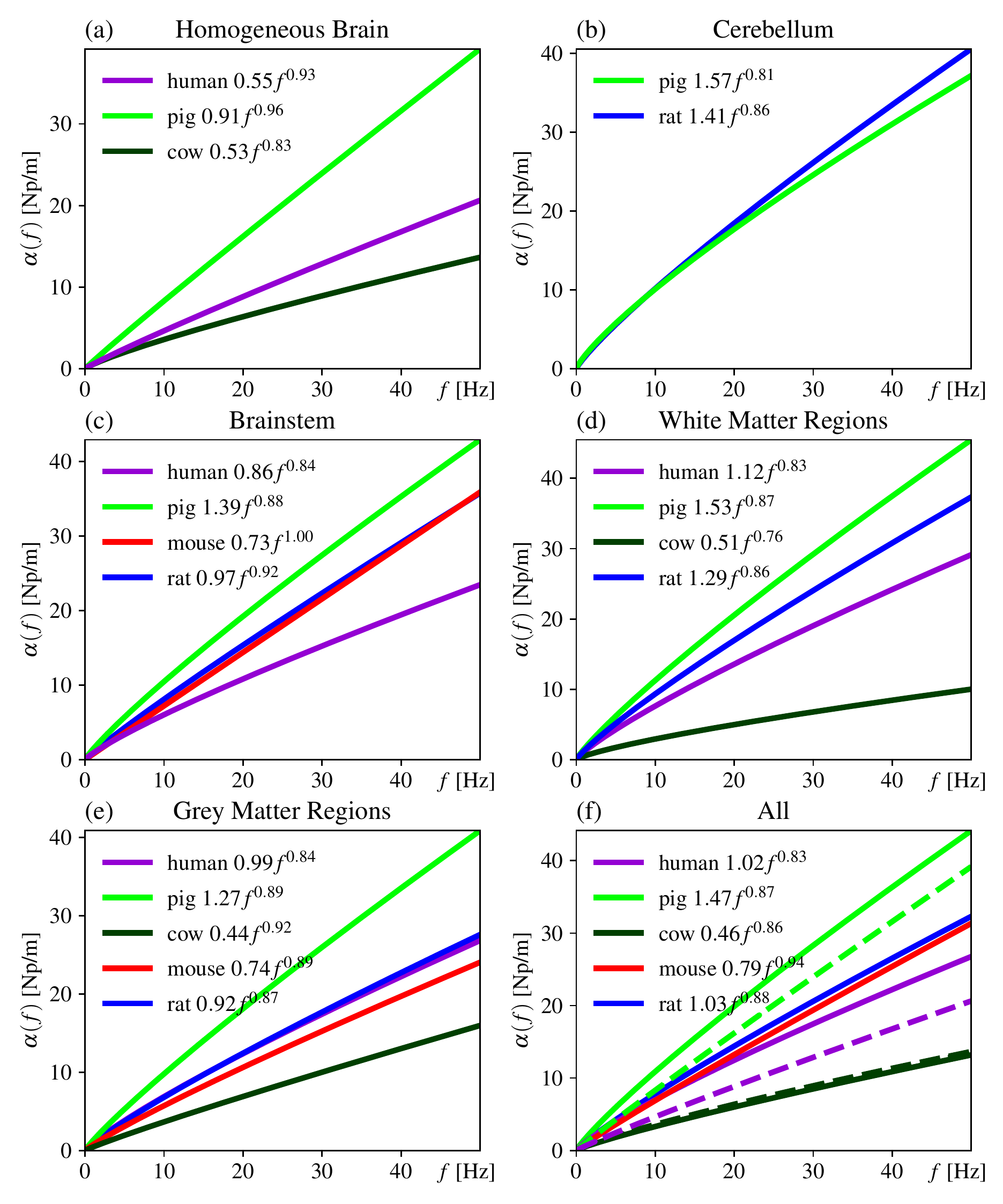}
    \caption{Averaged attenuation power laws in a low frequency regime (0-50 Hz) for a number of key regions, separated by animal type. {The ``all" fit is generated from merging the data for all regions bar the homogeneous brain (i.e. hippocampus, thalamus, brainstem, cerebellum, grey matter and white matter). Shown are the attenuation power law fits from experimental data for the different animal types for (a) the homogeneous brain, (b) cerebellum, (c) brainstem, (d) white matter, (e) grey matter and (f) ``all'' regions. Shown with a dashed line in subplot (f) are the fits from the homogeneous brain data as shown in subplot (a).}}
    \label{fig:animal}
\end{figure}

Experiments have already investigated this question, but only on a per-experiment basis. For example, MacManus \etal directly compared fits for human, pig, rat and mouse brains using indentation techniques \cite{MacManus2020}. 
Nicolle \etal similarly compared porcine tissue to human tissue using oscillatory experiments \cite{Nicolle2004}. 
Here we review across multiple experiments, specifically, we investigate the averaged properties from a large experimental literature segregated by animal type and region as shown in \autoref{fig:animal}.

{To achieve this, we employ the  same methodology as was done for the homogeneous brain characterisation used in FE models.}
For these comparisons, the following regions were considered: homogeneous brain, cerebellum, brainstem, white matter (i.e. corpus callosum and corona radiata data merged), grey matter (i.e. dentate gyrus, cortex and basal ganglia data merged) and ``all'' region created by merging data from all regions i.e. hippocampus, thalamus, brainstem, cerebellum, grey matter and white matter except the homogeneous brain. A total of eight different animals were found in our literature review - namely, pig, rat, human, mouse, cow, sheep, monkey and dog. However, due to scarcity of data from monkeys, sheep and dog, are not shown in \autoref{fig:animal}.

Our results found that the average power laws for each surrogate tissue do not always agree. {That is, the variation in results with respect to the use of different surrogate tissues is in fact significant.
This finding is not unexpected. For example, differences between human tissue and rodent brain tissue are anticipated since the rodent brain is quite anatomically different from the human brain \cite{Finan2012a}. Even for more anatomically similar tissues such as porcine and bovine tissue, differences are still observed in this work. We do however still note that there are other sources of variation due to different experimental techniques, post-mortem time, temperature, etc. that are also present in our dataset. For example, the data on human tissue comes primarily from indentation experiments \cite{Menichetti2020, MacManus2020, Sundaresh2022, Finan2017} whilst for bovine tissue it primarily comes from dynamic mechanical analysis \cite{Li2021}. Thus, we can anticipate that differences between the bovine dataset and human dataset will also occur due to differences in testing methods.
}

{We also point out some general trends observed here. 
We can see that the experimental data from human tissue is in fact generally less attenuating than porcine tissue, but more attenuating than bovine tissue, as can be observed for the homogeneous brain region (\autoref{fig:animal}a), white matter (\autoref{fig:animal}d), grey matter (\autoref{fig:animal}e), and the ``all'' region (\autoref{fig:animal}f). Rat and mouse tissues were found to be close.
Like in the previous section, here also we calculate the power laws for the ``all'' region and the homogeneous brain region are similar for the various animal types (\autoref{fig:animal}f). This is a promising finding since exact agreement is not expected - the data for the ``all'' region may be skewed towards various subregions depending on the data we have sourced. For example, 27\% of the data for rat comes from the cortex region.

It is particularly interesting to note that in this study the fits for larger animals such as porcine and bovine tissue were also found in general to be further from the fits for human tissue as compared to the fits for smaller animals such as rats and mice. This finding seems in direct conflict with the work of Dai \etal \cite{Dai2018}, who recommend the use of larger animals such as cows and pigs as surrogates over small animals such as rodents. However, there exists other work such as that of MacManus \etal \cite{MacManus2020} which suggest that mouse tissue is in fact a suitable surrogate. 

Furthermore, there are a number of reasons why we may observe this in this work. For one, it is important to keep in mind that there is relatively little data for rat and mouse brains, and the data that is presented lies in the low frequency regime (<10 Hz). As a result, the extrapolation to 50 Hz is exactly that: only an extrapolation. Thus, any comparisons at higher frequencies should be done carefully. Furthermore, we also point out that the cortex region is also the most commonly experimented upon tissue for the rat, mouse and human data, whilst this is not the case for the porcine and bovine tissues. These differences in sampled subregions of the brain may also partly explain the trends observed in this work.
{However, this still does not fully explain why the porcine tissue seems to be substantially more attenuating than other tissue types. This phenomenon instead appears to come due to different experimental techniques. In the collected literature, the dominant experimental technique for porcine tissue is indentation tests \cite{Elkin2011b, MacManus2017a, MacManus2017c, Sundaresh2021, Pan2019, Pan2022}. This is also a common experimental technique for other surrogates such as rat also, but the experimental results for porcine tissues are substantially different. Specifically, there is a disproportionate amount of experiments on porcine tissue in the literature that find low instantaneous shear moduli, which leads to high predictions for the attenuation. For example, of all the Prony-series data collected for porcine tissue, 55\% of the pig data has an instantaneous shear modulus less than 1500 Pa. By contrast, for rat tissue it is merely 23\% and for human it is 33\%. This difference does not appear amongst experiments which have conducted experiments on both porcine and other surrogates using the same experimental procedure \cite{MacManus2020, Nicolle2004}. Instead, this arises from the fact that there are experiments conducted \emph{solely} on porcine tissue which report low values for the instantaneous shear modulus \cite{Ning2006, Sundaresh2021, Miller1997, Miller1999, Miller2002, Gefen2004, Prange2000}.}
In particular, this finding emphasises that the use substitute data from surrogate tissues must be done with much caution.}

\subsection{Limitations and shortcomings}

This study does have some limitations. First, the reference values were not varied per region in this work and we took $\rho=1000$ kg/m$^3$ and $c=2.1$ m/s at 75 Hz for all regions and has been fixed for $M_0$ calculations. 
The use of a constant density is in line with the approach of FE models but is nonetheless limiting. 
The reference dispersion value is obtained with the assumption of homogeneous brain tissue. 
Thus, it may not be suitable for tissues that are very different from the homogeneous brain like meninges and spine. 
We were unable to include these tissues as a result, although there does exist experimental Prony-series data for them (see \cite{Ramo2018a, Ramo2018b, Shetye2014, Jannesar2018, Bass2007, Troyer2012a, Troyer2012b}).

Another key issue is the variations of the experimental datasets in the literature.
Since there are many possible sources of variation and it is not feasible to account for all of them at the present time, particularly given that the literature does not even always agree on their effects. 
For example, there is some dispute about whether or not there exists a sex-dependence of brain tissue properties \cite{Budday2020}. 
It is hoped that by averaging across many series in this work, the variations will even out to some degree.

{Lastly, the reliance on Prony-series data is also limiting, particularly when considering frequency-domain quantities. As mentioned previously, the use of a limited number of mechanisms in a Prony-series causes oscillatory artefacts to appear in the predicted inverse quality \cite{Blanc2016, Emmerich1987}. Thus, it would be better to directly use data from the frequency domain for such quantities, but this is not what current FE models are predominantly doing. Furthermore, during the literature it was found that more experimental papers yielding Prony-series data were available as compared to frequency-domain data.}
As a result, this was a necessary limitation to introduce to this work. 
Similarly, the curve fitting of a Prony-series is also limiting but necessary in order to give results that can be used by FE models. 
Nonetheless, we also provide the direct power law fit so frequency-domain data is also available in this work. 
It is also important to stress that curve fitting for $Q^{-1}$ is not a trivial exercise and differing fits can be possible for the same data depending on the algorithm and initial conditions used \cite{Fung1993}. 
Thus, we recommend that users conduct their own curve fits which they can tailor specifically to their application. In this work, the power laws are provided to facilitate this.
\subsection{Recommendations for future work}

The dominant method of modelling viscoelasticity for current state-of-the-art FE models is by means of a Prony-series, though some models, such as the LiUHead model \cite{LIUHEAD}, have opted for other approaches. This state of affair is unlikely to change in the immediate future, but there are a number of improvements we can suggest to current techniques. 

First, many FE models are incorporating viscoelasticity by means of a one-term Prony-series \cite{Tse2014}, which greatly limits the frequency range that can be modelled, especially if one is interested in modelling the transient visco-elastic behaviour. There also exists a large range of higher order viscoelastic models in the literature which are included in this work and these laws could be leveraged instead.

Furthermore, this work and many others \cite{Antonovaite2021, Hiscox2020, Budday2020} have established that the brain is heterogeneous, whilst it is often times treated as a homogeneous tissue. In some cases, properties for certain tissues as used in FE models have also been derived from experiments on different tissues - for example, the homogeneous brain properties of the ADAPT \cite{ADAPT}, ANISO KTH v1 \cite{ANISOKTHv1}, ANISO KTH v2 \cite{ANISOKTHv2}, WHIM v1 \cite{WHIMv1}, ICM \cite{ICM} and KTH v2 \cite{KTHv2} models are taken from experiments by Nicolle \etal \cite{Nicolle2005} on corona radiata tissue. Similarly, the homogeneous brain properties derived from the experiments of Shuck \& Advani \cite{Shuck1972} are also derived from corona radiata tissue.
This is could lead to erroneous results and should be used with caution.

There are also differences between the tissues chosen for inclusion in FE models versus the tissues that are experimented on. 
Experiments can provide different measurements for specific regions compared to the larger regions taken by FE models. For example, the cortex region which is sometimes included in FE models is measured in a total of six subregions by Menichetti \etal \cite{Menichetti2020} - namely the prefrontal cortex, posterior-occipital cortex, superior mid-frontal cortex, postero-lateral frontal cortex, inferior temporal cortex and the postero-superior frontal cortex. 
A clear difference is also that FE models are currently often modelling the two regions of white and grey matter whilst typically experimental papers are not. Instead, a significant amount of experimental work measures subregions of these - white matter is commonly measured as either the corona radiata or the corpus callosum, and grey matter as the basal ganglia, cortex or the dentate gyrus. Furthermore, our work has found that these subregions are mechanically different (see \autoref{fig:low_freq_atten}). Thus, these subregions should be considered separately in future work.

We also point out that taking viscoelastic and hyperelastic data from different experiments can be problematic as viscoelastic fits can change depending on the hyperelastic model used, and also vary in general between experiments. 
Our work also shows that experimental data and data used in FE models agree with each other. Thus, we recommend using directly experimental measurements in future work as opposed to modifying or scaling experimental data. 
In this work we provide both averaged laws for twelve regions and eight different animals and also a total of 181 different Prony-series in order to facilitate this.

Lastly, as was mentioned in the limitations section, the use of Prony-series is not ideal. Future work could directly obtain averaged laws from frequency-domain data i.e. values of $M'$ and $M''$.

\section{Conclusion}

To the best of our knowledge, this work presents 1) the first multi-frequency viscoelastic atlas of the heterogeneous brain, 
2) the first review focusing on \emph{viscoelastic} modelling in \emph{both} FE models and in experimental works, 3) the first attempt to conglomerate the disparate existing literature on the viscoelastic modelling of the brain. Thus, our review differs from existing work in a number of key ways.

Our review enables a direct comparison between the experimental literature and the data used in FE models. Existing reviews focus typically on either reviewing FE models, or reviewing experimental techniques, but not both together. This review aims to help bridge the gap between these two domains. To this end, we have gathered a total of 181 differing Prony-series from 48 different experimental papers and 31 unique Prony-series used in FE models. This review gives the largest collection of viscoelastic parameters for human brain tissue. This wealth of data allows us to investigate differences due to animal tissue choices in the heterogeneous brain with greater granularity, for instance, we can now compare corona radiata of a  pig brain with that of the cortex of the human brain unlike previous studies. Our work also provides a means of comparing Prony-series viscoelasticity to storage and loss moduli data (e.g. from MRE measurements), and to attenuation laws. 
{Previous works have not thoroughly investigated the link between relaxation functions and storage and loss moduli. For example, the review of Chatelin \etal \cite{Chatelin2010} provides many different experimental results for relaxation functions, and also many different distinct experimental results for storage and loss moduli. However, their review does not investigate how the predictions of the storage and loss moduli from the relaxation function $g(t)$ compare with the other experimental data {for the storage and loss moduli}.}
 
Comparison of FE model data with the recent experimental data yields that FE models are generally underestimating the attenuation than the recent experimental data. {Our review uncovers that there may be issues with existing commonly used Prony-series data. For example, the most used dataset in FE models is the one presented by Shuck and Advani \cite{Shuck1972}. However, their data is much stiffer than the average calculated using our approach. They have found that for a frequency range $[3, 300]$ Hz, the storage modulus $M' \in [7, 30]$ kPa and loss modulus $M'' \in [1, 90]$ kPa, whilst in this work our average Prony-series predicts lower values for both the storage modulus $M' \in [1,6]$ kPa and loss modulus $M'' \in [0.3, 1]$ kPa. It is thus clear from both this work and other previous reviews such as Chatelin \etal \cite{Chatelin2010} and Hrapko \etal \cite{Hrapko2008} that the data of Shuck and Advani is an outlier with respect to the rest of the experimental literature. In addition, another commonly used dataset, namely that of Nicolle \etal \cite{Nicolle2004} was found to predict $Q^{-1}\geq1$.} Therefore there is a need to recalibrate and reassess the material properties used in the computational models describing the brain trauma.
 
{We calculate the average attenuation power law for the homogeneous brain tissue from recent experimental data {(obtained from 18 unique Prony-series)} as $\alpha(f)=0.70f^{0.91}$ Np/m.}
The corresponding average dimensionless Prony-series is $\hat{g}(t) = 0.0872 + 0.0722e^{-0.0385t} +
0.122e^{-0.844t} + 0.22e^{-17.6t} + 0.498e^{-396t}$, with an instantaneous shear modulus of $M_0 = 5167$ Pa at 75 Hz. {The mean and median instantaneous shear modulus in the experimental literature are 6230 Pa and 3750 Pa, respectively.}

{Significant differences are also observed between the animal types, with relative errors \\ $\displaystyle \frac{\left( a_{\rm Human}\omega^{b_{\rm Human}} - a_{\rm Surrogate}\omega^{b_{\rm Surrogate}}\right)}{a_{\rm Human}\omega^{b_{\rm Human}}}\times 100$ of 23-38\% for bovine tissue and 78-95\% for porcine tissue for the attenuation power law fits between 10 and 100 Hz for the homogeneous brain region. This emphasises the need to take caution when using surrogate tissues, since substantial differences can exist. }

In addition, this work provides a methodology for computing the predictions of a given Prony-series on the storage and loss moduli, quality factor, dispersion relation and attenuation. {Since we have been able to calculate averaged Prony-series and power laws, it also provides a useful methodology for investigating and comparing an experimentally obtained Prony-series to the rest of the experimental literature. 
Importantly, it is also possible to verify whether or not a Prony-series predicts an inverse quality satisfying $Q^{-1}(\omega) < 1$ for it to be physically viable. Thus Prony-series which do not satisfy this may need to be recalibrated.}
}
From a numerical stand point, the methods using one- or two- term Prony-series can limit the attenuation and dispersion modelling especially in the nonlinear regime which results in the generation of higher harmonics such as shear shock formation in brain \cite{Tripathi2021}. 

\clearpage
\appendix

\section{Appendix: Validation of averaged viscoelastic properties}

\begin{figure}[h!]
    \centering
    \includegraphics[trim=10 10 1 10, clip, width=0.8\textwidth]{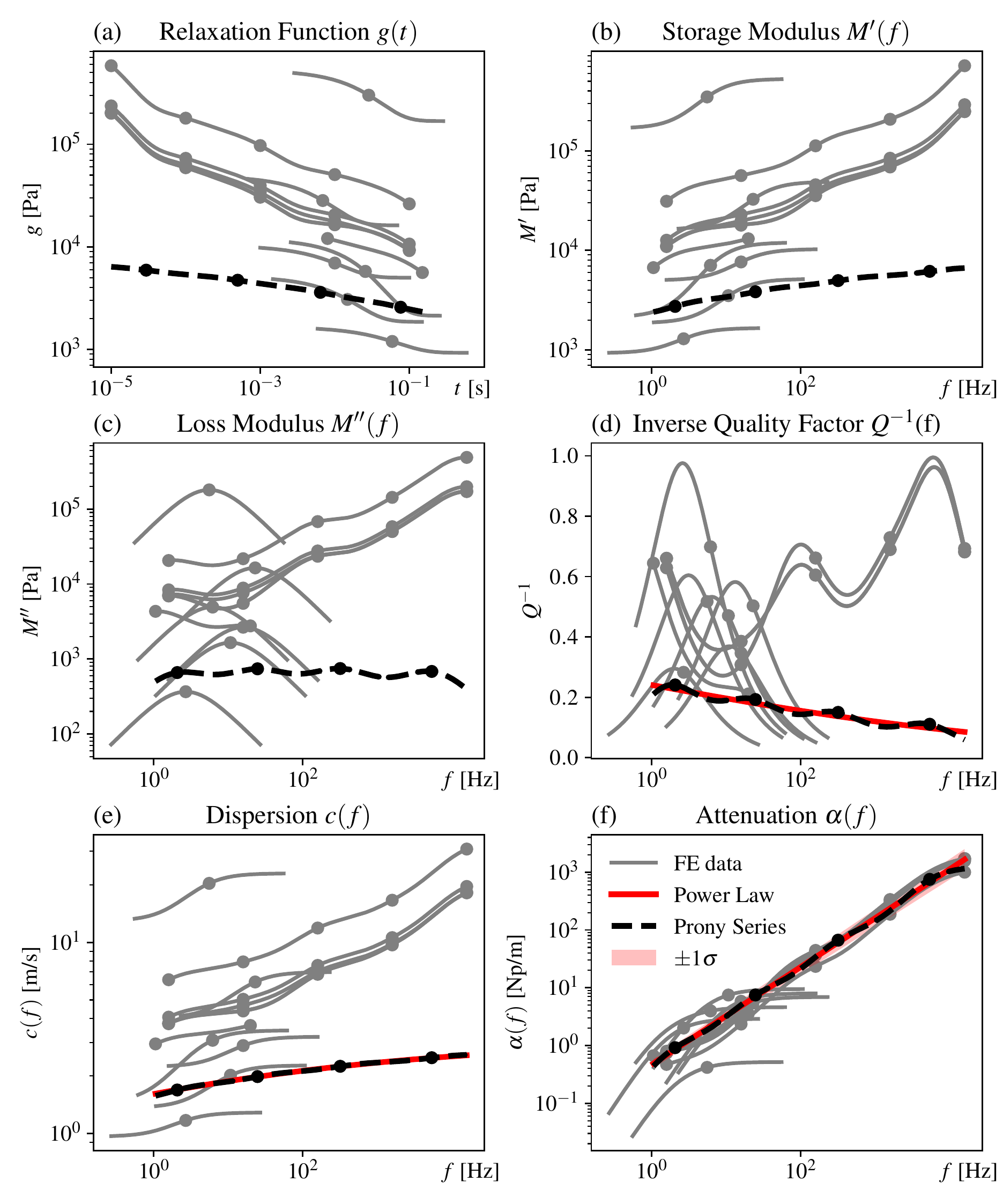}
    \caption{Derivation of averaged attenuation power laws, and corresponding dispersion and quality (shown in red). An averaged Prony-series and its predictions are also shown in black. {Shown are the predictions for (a) the relaxation function, (b) storage modulus, (c) loss modulus, (d) inverse quality factor, (e) dispersion and finally (f) attenuation.}}
    \label{fig:fem_brain_fitted}
\end{figure}

{Our procedure for determining averaged Prony-series data involves a number of nontrivial steps and thus it is important to verify that our method proceeds as expected. Specifically, a number of sensible checks can be conducted:
\begin{itemize}
    \item Can our averaged Prony-series reconstruct the averaged power law from which it was derived?
    \item Does our averaged Prony-series lie amongst the experimental data from which it was derived?
    \item Does our averaged Prony-series or attenuation power law predict $Q^{-1} < 1$ as expected?
    \item Does our determined value of $M_0$ from our Prony-series match that of the experimental data?
\end{itemize}

We illustrate this procedure for the experimental data on the homogeneous brain tissue. Details for other tissue types can be found in the supplementary materials.

Following the obtention of an averaged Prony-series as shown in \autoref{fig:workflow}d, the forward calculation step can be conducted on this new Prony-series as outlined in 
\autoref{fig:workflow}e. The results of this process are shown in \autoref{fig:fem_brain_fitted}. 

As evident from \autoref{fig:fem_brain_fitted}a, the averaged Prony-series  (dashed black) calculated using $c(\omega)$ given the Kramers-Kronig relations underestimates the storage/loss modulus and the inverse quality (\autoref{fig:fem_brain_fitted}b-e, respectively). However, it is able to reconstruct the attenuation power law (red) shown in \autoref{fig:fem_brain_fitted}f.
The underestimation of the storage/loss modulus and the inverse quality is due to the use of Kramers-Kronig relations \cite{Waters2000} which may not be ideal for the point estimates  provided for Prony-series, moreover the use of the reference value of $c=2.1$ m/s at 75 Hz further restricts the approximation. However, it does provide a benchmark to unify the different observations obtained using different experimental techniques. 
}

{Lastly, we can also examine the prediction for $M_0$ from our averaged Prony-series, since this an important experimental quantity in the literature. It is important to determine whether or not the prediction from equation \eqref{eq:get_M_0} is in line with the distribution of values of $M_0$ from the literature. In general, quite a lot of variation exists in the predictions for the instantaneous shear modulus since this can depend upon experimental techniques and procedures. It is not possible to experimentally measure a value for the relaxation function at $t=0$, so differing values of $M_0$ can occur depending on what time interval (or frequency range) one investigates. In our work, we find our averaged Prony-series for the experimental homogeneous brain tissue has a value $M_0=5167$ Pa at 75 Hz. This broadly agrees with the experimental literature, which has a mean value of 6230 Pa and a median value of 3750 Pa. }

\clearpage
{\footnotesize 

}

\end{document}